\documentclass[11pt,british]{article}
\usepackage[utf8]{inputenc}
\usepackage{mathtools}
\usepackage{amsmath}
\usepackage{babel}
\usepackage{pdflscape}
\usepackage{amsthm}
\usepackage{amssymb}
\usepackage{setspace}
\usepackage{microtype}
\usepackage{slashed}
\usepackage{cancel}
\usepackage{multirow}
\usepackage{bbm}
\usepackage{booktabs}
\usepackage{subcaption}
\usepackage{longtable}

\usepackage{placeins}

\usepackage{pgfplots}
\pgfplotsset{compat=newest}
\usepgfplotslibrary{fillbetween}

\usepackage{import}

\makeatletter
\usepackage{tikz}
\usepackage{tikz-cd}
\usetikzlibrary{3d} 
\usetikzlibrary{backgrounds}
\usepackage[framemethod=tikz]{mdframed}
\AtBeginEnvironment{mdframed}{%
\tikzset{every picture/.style={}}%
}
\mdfsetup{roundcorner=.5ex}

\theoremstyle{definition}
\newtheorem*{defn*}{Definition}

\usepackage{jheppub}                   
\usepackage[linecolor=blue,backgroundcolor=blue!25,bordercolor=blue,textsize=scriptsize]{todonotes}

\makeatletter
\gdef\@fpheader{\ }                    
\makeatother

\usepackage{slashed}	 	
\usepackage{amsfonts} 		
\SetTracking{encoding={*}, shape=sc}{0} 	
\usepackage{color} 		
\definecolor{darkblue}{rgb}{0.0,0.0,0.3} 	
\allowdisplaybreaks		
\date{\today} 		
\numberwithin{equation}{section}	

\makeatletter
\g@addto@macro\bfseries{\boldmath}
\makeatother

\let\originalleft\left
\let\originalright\right
\renewcommand{\left}{\mathopen{}\mathclose\bgroup\originalleft}
\renewcommand{\right}{\aftergroup\egroup\originalright}

\definecolor{darkgreen}{RGB}{53,199,50}

\newcommand{\SO}[1]{\mathrm{SO}(#1)}
\newcommand{\SU}[1]{\mathrm{SU}(#1)}
\newcommand{\Orth}[1]{\mathrm{O}(#1)}

\newcommand{\bbR}{\mathbb{R}}

\newcommand{\bbZ}{\mathbb{Z}}
\newcommand{\bbG}{\mathbb{G}}
\newcommand{\bbD}{\mathbb{D}}
\newcommand{\dd}{\mathrm{d}}

\DeclareMathOperator{\tr}{tr}
\DeclareMathOperator{\Tr}{Tr}
\DeclareMathOperator{\diag}{diag}

\title{Machine Learning the 6d Supergravity Landscape}
\author[a]{Nathan Brady,}
\emailAdd{bradyns@tamu.edu}
\author[a]{David Tennyson}
\emailAdd{dtennyson@tamu.edu}
\author[a]{and Thomas Vandermeulen}
\emailAdd{tvand@alum.mit.edu}
	
\affiliation[a]{Mitchell Institute for Fundamental Physics and Astronomy, Texas A\&M University, College Station, TX, 77843, USA}

\abstract{In this paper, we apply both supervised and unsupervised machine learning algorithms to the study of the string landscape and swampland in 6-dimensions. Our data are the (almost) anomaly-free 6-dimensional $\mathcal{N} = (1,0)$ supergravity models, characterised by the Gram matrix of anomaly coefficients. Our work demonstrates the ability of machine learning algorithms to efficiently learn highly complex features of the landscape and swampland. 
Employing an autoencoder for unsupervised learning, we provide an auto-classification of these models by compressing the Gram matrix data to 2-dimensions. Through compression, similar models cluster together, and we identify prominent features of these clusters. The autoencoder also identifies outlier models which are difficult to reconstruct. One of these outliers proves to be incredibly difficult to combine with other models such that the $\tr R^{4}$ anomaly vanishes, making its presence in the landscape extremely rare.
Further, we utilise supervised learning to build two classifiers predicting (1) model consistency under probe string insertion (precision: 0.78, predicting consistency for 214,837 models with reasonable certainty) and (2) inconsistency under anomaly inflow (precision: 0.91, predicting inconsistency for 1,909,359 models). Notably, projecting these predictions onto the autoencoder's 2-dimensional latent layer shows consistent models clustering together, further indicating that the autoencoder has learnt interesting and complex features of the set of models and potentially offers a novel approach to mapping the landscape and swampland of 6-dimensional supergravity theories.}

\subheader{\hfill\textrm{MI-HET-856}}

\begin{document}

\maketitle

\section{Introduction}

In recent years, the swampland program started in \cite{Vafa:2005ui} has determined numerous constraints that effective theories must satisfy in order to be able to consistently couple to a quantum theory of gravity (see e.g. \cite{vanBeest:2021lhn} and references therein for a review). Instead of trying to derive results directly from string theory, the swampland conjectures are derived in a model independent manner, from universal physical properties that \emph{any} consistent quantum theory of gravity should satisfy. These are often tested in string theory, then applied to a variety of effective quantum field theories (QFTs) to try to determine the `landscape' (models which can be coupled to quantum gravity) from the `swampland' (those which can't). This method has seen great success and has, for example, shown that all consistent theories of supergravity in dimension 7 or greater can be obtained from string theory \cite{Hamada:2021bbz,Montero:2020icj,Bedroya:2021fbu}.

These results were helped, in part, by the relatively limited number of anomaly-free models that can be written down with 16 supercharges. For example, it is known that the rank of the gauge group in a $d$-dimensional supergravity theory with 16 supercharges is bound by $r\leq 26-d$. This allowed for direct computation and comparison with swampland conjectures in each case explicitly. The situation in 6-dimensional $\mathcal{N}=(1,0)$ supergravities, the highest dimension with 8 supercharges, is drastically different. Local anomaly freedom and the absence of ghosts allows for an infinite number of models \cite{Kumar:2010ru}. Other swampland criteria based on absence of global anomalies \cite{Monnier:2018cfa,Monnier:2018nfs,Davighi:2020kok}, constraints on the Green-Schwarz anomaly cancelling term and the charge lattice \cite{Monnier:2017oqd}, and consistency when coupled to probe branes \cite{Kim:2019vuc} have put constraints on the possible gauge groups and representation content \cite{Kim:2019vuc,Lee:2019skh,Angelantonj:2020pyr,Cheng:2021zjh,Tarazi:2021duw}. Despite this, there still remain infinite families which pass all known consistency criteria \cite{PhysRevD.109.126006}. Even excluding these infinite families, millions of models have been explicitly tabulated \cite{Avramis:2005hc,Hamada:2023zol,Becker:2023zyb}. It seems that, lacking some new swampland criteria which significantly restricts the possible consistent models in 6-dimensions, direct analysis of each of these models is not possible. Instead, the wealth of examples of seemingly consistent theories may indicate that a statistical approach to analysis may be fruitful.

Machine learning is an approach to analysing data, where one models some complicated function by a neural net (see e.g. \cite{LeCun2015} and references therein, or \cite{Ruehle:2020jrk} for a pedagogical review with applications to string theory). At its simplest, a neural net (NN) consists of various layers of operations which take as input a vector $\mathbf{n}_{i}\in \bbR^{k_{i}}$ and returns $\mathbf{n}_{i+1}=\sigma(A_{i}\mathbf{n}_{i} + \mathbf{b}_{i})$, where $A_{i}$ is a $k_{i+1}\times k_{i}$ matrix, $\mathbf{b}_{i} \in  \bbR^{k_{i+1}}$, and $\sigma$ is some non-linear activation function applied componentwise. The vector $\mathbf{n}_{i+1}$ is used as input for the next layer. In the final layer, the output is compared with some desired result, and the error is measured by some loss function $L$. By passing multiple data points (called the training data) through the neural net, the machine learning algorithm tunes the parameters $A_{i},\mathbf{b}_{i}$ to minimise the loss function with respect to the seen data. There are other parameters, known as hyperparameters, which can be chosen to try to reduce the loss further. These include changing the activation function $\sigma$ and the neural net architecture, referring to the shape of the layers and the precise way in which they are connected. Given enough training data which is representative of the population and a well-designed architecture, the neural net should provide a good approximation to the complicated function and can be applied to new, unseen data. In principle, given a large enough neural net, one can approximate any (continuous bounded) function to an arbitrarily high accuracy \cite{Cybenko1989}. In practice, however, there are significant computational limitations and a reasonable amount of trial and error is required to get an accurate result from the machine learning algorithm.

The process of minimising a loss function has made machine learning appealing for many physical applications. For example, NN's have been used to find numerical approximations to Calabi-Yau metrics by taking points on the variety as input, and the metric components or K\"ahler potential as output \cite{Douglas:2020hpv,Halverson:2023ndu,Jejjala:2020wcc,Ashmore:2019wzb,Ashmore:2021ohf,Larfors:2021pbb}. They have also been used to find ground states of lattice gauge theories by using the NN to approximate the wavefunction \cite{Apte:2024vwn}. In both cases, one utilises the fact that there is a functional, which is minimised by the desired function, which can play the role of the loss function. In the former case, the loss is given in terms of some parameter $\eta$ which is the ratio of volumes defined by $2^{-n}\Omega\wedge \bar{\Omega}$ and $(n!)^{-1}\omega^{n}$. The Ricci flat metric occurs precisely when $\eta=1$, so the loss is measured by the integral of $(1-\eta)^{2}$. The learning procedure effectively implements a form of Ricci flow on the manifold. In the latter case, the Hamiltonian plays the role of the loss function and Monte Carlo methods are used to approximate gradients. Similar techniques have also been used to find Hermitian-Yang-Mills connections for heterotic backgrounds \cite{Ashmore:2021rlc}. It was also shown in \cite{Cipriani:2025ini} that physics-informed NN's can outperform standard integrators for finding geodesics of a massless particle moving within critical regions of a D1-D5 circular fuzzball geometry.

More conventionally, machine learning algorithms fall into 3 rough categories - supervised, unsupervised, or reinforcement learning. In supervised learning, one provides labelled data which associates with each input datum an accurate label which we want the algorithm to predict. Such models are good for classification problems or regression analysis. In physics, these techniques have been used to predict properties of Calabi-Yau manifolds relevant for physical applications. These properties include Hodge numbers \cite{He:2017aed,Bull:2019cij,Bull:2018uow}, whether the Calabi-Yau is elliptically fibred \cite{He:2019vsj}, and their volumes \cite{Krefl:2017yox}. Supervised learning has been used to accurately predict the gauge groups appearing in F-theory constructions \cite{Wang:2018rkk,Carifio:2017bov}. Finally, supervised learning has been shown to be able to distinguish between `standard-model-like' and `non-standard-model-like' models of heterotic line bundle models \cite{Deen:2020dlf}. 

Unsupervised learning, by contrast, encompasses a variety of techniques which use unlabelled data in the training process. Since the data is unlabelled, the algorithm is not told which features to learn, and instead identifies prominent features on its own. These techniques can be useful for pattern recognition, and data compression. One particularly important technique, called an autoencoder (to be reviewed later), can do both of these by compressing an input vector down to a lower-dimensional latent space. By passing enough data through the autoencoder, patterns can be recognised through clustering of points in the latent layer \cite{Hinton2006}. Autoencoders were used in \cite{Deen:2020dlf}, where it was shown that standard-model-like and non-standard-model-like solutions clustered separately in the latent layer. Different unsupervised clustering methods were used in \cite{Klaewer:2018sfl} to help identify analytic formulae for line bundle cohomologies on hypersurfaces in toric varieties. In \cite{Mutter:2018sra}, autoencoders were used to search for realistic models in the space of heterotic orbifold theories.

Reinforcement learning is different to the other two learning algorithms in that the algorithm is not presented with representative data. Instead, the algorithm randomly searches through the solution space, and is provided with a success score based on some measure of success. The score is fed back to the algorithm to help refine the probability distribution defining the random search. Such techniques will not be the focus of this paper, but we point to \cite{Kantor:2021kbx,Kantor:2021jpz,Kantor:2022epi} for interesting applications to bootstrapping SCFTs, and to \cite{Halverson:2019tkf,Ruehle:2017mzq} for reinforcement learning for finding string vacua and studying the string landscape. Similar approaches using various evolutionary algorithms have also been used in \cite{Damian:2013dq,Damian:2013dwa,CaboBizet:2020cse,Cole:2019enn,Abel:2014xta,Bena:2021wyr,AbdusSalam:2020ywo} to explore the string landscape and flux compactifications.

In this paper we perform an initial study of the application of both supervised and unsupervised learning to gauged supergravities in 6 dimensions. As a training set of data, we use the roughly 26 million models identified in \cite{Hamada:2023zol} as building blocks of anomaly-free 6-dimensional supergravity models. They are building blocks in the sense that they satisfy all local anomaly criteria\footnote{And global criteria for $G_{2}$ \cite{Bershadsky:1997sb}} except for the gravitational anomaly. However, they can be combined in a trivial way with another model in the set, i.e. with no cross-representation content of the relevant gauge groups, in order to build a completely anomaly-free model. As input for our NN models, we take the upper triangle of the Gram matrix (i.e. the matrix of anomaly coefficients assuming the $\tr R^{4}$ and $\tr F^{4}$ terms have vanished), flattened into a 136 dimensional vector.

For the unsupervised algorithm, we train a feed-forward autoencoder which compresses the Gram matrix representation into a lower dimensional space (the latent space). We find that we can compress the data to two dimensions with low reconstruction loss, and hence we can plot the output of the latent layer. We find interesting clustering which we explore in detail in \cite{ml-github-website}, and which may provide a new way to classify models. Further, by looking at the models with the highest reconstruction loss, we use the auto-encoder to automatically look for models which are peculiar compared to the majority. This method of identifying peculiar data points is known as `anomaly detection' in the machine-learning literature, but we refer to it as `peculiarity detection' to avoid confusion with the more common notion of anomaly in high energy theory. In doing so, we efficiently identify a model which, from the representation content and gauge groups does not seem particularly special, but upon further examination proves to be very difficult to combine with other models into a fully anomaly-free theory. It is quite remarkable that the machine learning algorithm should be able to identify such a model since it was only given the Gram matrix with which to learn. It was not given, for example, any information on how to combine theories.

For the supervised learning algorithm, we train a classifier on a subset of the models in \cite{Hamada:2023zol}, with the aim of identifying models which do, and do not, pass the anomaly inflow criteria of \cite{Kim:2019vuc}. To label the data, we develop a numerical method to check whether the inequalities required for unitary probe branes are satisfied. We develop two labelled datasets - one which identifies models where the inequality is satisfied everywhere in a region around the origin, and another where the inequality is not satisfied in a region around the origin - and hence train two classifiers. While the first set satisfy a necessary condition (but not sufficient) for freedom from inflow anomalies, the second condition is neither necessary nor sufficient but is a strong constraint on the model. Hence, the first classifier can be viewed as identifying models which pass a (restricted) anomaly inflow criteria, while the second identifies models which are likely to be inconsistent. After training the classifiers on the subset of the data we obtain a precision of 0.78257 for the first classifier, and 0.90933 for the second. 

Note that, while our method is not exact, it can analyse the roughly 26 million models very efficiently. By applying it to the full set, we identify over 214,000 models from the first classifier (classifier-0) which would likely pass the inflow criteria, and 1,900,000 from the second classifier (classifier-1) which are likely inconsistent. Making such predictions by hand would be almost impossible, and discrete optimisation methods for exact analysis of of the inflow anomaly inequalities are computationally expensive. Instead, while not a proof, our work efficiently identifies both a fertile landscape of building blocks (output of classifier-0) which can be used to search for further anomaly-free models, and a potential swampland of building blocks which would lead to inconsistent models when combined with any other building block. Finally, we plot the identified consistent and likely inconsistent models in the latent space of the autoencoder and find that the output of the first classifier form clusters. This further evidences the ability of the autoencoder to identify highly complex properties of the models from just their Gram matrices.

The paper is laid out as follows. In section~\ref{sec:gsugra}, we review supergravity theories in 6-dimensions and discuss the consistency conditions arising from local and global anomaly freedom, ghost freedom, consistency conditions on the charge lattice, and freedom from inflow anomalies. We also review the format of the $\sim$26 million models defined in \cite{Hamada:2023zol} which serve as our dataset.  Section~\ref{ML-Auto} then trains an autoencoder on this data, which provides both an automatic means of classification through clustering analysis and a way to detect models with peculiar features.  In Section~\ref{sec:numerical_labelling} we pivot to supervised learning in the form of a classifier which we use to attempt to identify models which satisfy all known consistency conditions.  Finally, Section~\ref{sec:summary} concludes with a discussion of possible future directions. Appendix \ref{app:A} covers an ambiguity in the anomaly inflow criteria for models with sufficiently degenerate Gram matrix, which we must account for in the numerical method for labelling the data.

All of the data we use in this paper, the trained neural nets, and their output can be found in the GitHub repository \cite{ml-github-repo}. The clustering is detailed in the accompanying website \cite{ml-github-website}. The original raw data was taken from the GitHub repositories \cite{GLogesData1,GLogesData2} which accompany the work of \cite{Hamada:2023zol}.

\section{\texorpdfstring{$\mathcal{N}=(1,0)$}{N = (1,0)} supergravity in 6 dimensions}
\label{sec:gsugra}

The $\mathcal{N}=(1,0)$ supersymmetry algebra in 6-dimensions allows for the following multiplets\footnote{We shall follow the notations and conventions of \cite{BST} throughout this section.}.
%
\begin{equation}
    \underbrace{ \left(e_\mu^m, \psi_{\mu -}^A, B_{\mu\nu}^{+} \right)}_{\text{graviton}}\ ,\qquad \underbrace{\left(B_{\mu\nu}^{-}, \chi_{+}^A,\varphi  \right)}_{\text{tensor}}\ ,\qquad \underbrace{\left(A_\mu, \lambda_{-}^A\right)}_{\text{vector}}\ ,\qquad \underbrace{\left(4\phi,\psi_{+} \right)}_{\text{hyper}}\ .
\end{equation}
The 2-form potentials $B^{\pm}$ have (anti)-self-dual field strengths. The fermions are symplectic Majorana-Weyl, with $A$ labelling the doublet of the $\mathrm{Sp}(1)$ R-symmetry group, and their chirality denoted by $\pm$. To specify a gravitational theory, it is enough to specify the number $n_{T}$ of tensor multiplets, the gauge group $G$ which determines the vector multiplets, and a set of representation content $R$ for the hypermultiplets. The couplings in the theory are determined by supersymmetry and anomaly cancellation. We will often denote the total number of vector and hypermultiplets as $n_{V} = \dim G$ and $n_H = \dim R$ respectively. There are many consistency requirements on 6-dimensional supergravity theories arising from anomaly cancellation and swampland criteria which we review now.

\subsection{Review of consistency conditions}\label{sec:consistency_review}

\subsubsection*{Local anomaly freedom}

Since the theory is chiral, it may be anomalous, with possible gravitational anomalies arising from the gravitino and 2-form fields, and further gravitational, gauge or mixed anomalies arising from the gaugino and hyper-fermions. The local anomalies can be determined from an 8-form anomaly polynomial which is, in turn, determined by the matter content of the theory. In this paper, we will be interested in models where the gauge group is semi-simple and no subgroup of the R-symmetry group is gauged. We will write the gauge group and representation content as
\begin{equation}
    G = G_{1}\times ... \times G_{n} \ , \qquad R = R_{1}\otimes ...\otimes R_{n}\ ,
\end{equation}
where each $G_{i}$ is simple and the $R_{i}$ denote (possibly reducible) finite dimensional representations of the $G_{i}$. The contribution to the anomaly polynomial from each of the fields is as follows 

\begin{align}
    P(\psi_{\mu -}) &= c\Big( -\frac{245}{360} \tr R^4 +\frac{43}{288} (\tr R^2)^2\Big) \ ,\\[5pt]
    P(\chi_+) &= c n_T \Big( \frac{1}{360} \tr R^4 + \frac{1}{288} (\tr R^2)^2 \Big)\ ,\\[5pt]
    P(B_{\mu\nu}) &= c (n_T-1) \Big( \frac{28}{360} \tr R^4 - \frac{1}{36} (\tr R^2)^2 \Big)\ ,\\[5pt]
    P(\lambda_-) &= c \Big\{ -n_V  \Big[ \frac{1}{360} \tr R^4 +\frac{1}{288} (\tr R^2)^2\Big]
    +\frac{1}{6}  \tr R^2   \sum_{i=1}^n \Tr F_i^2 
    -\frac{2}{3}  \sum_{i=1}^n \Tr F_i^4   \Big\}\ ,\\[5pt]
    P(\psi_+) &= c\Big\{ n_H \Big[ \frac{1}{360} \tr R^4 +\frac{1}{288} (\tr R^2)^2 \Big] 
     - \sum_{i=1}^n \sum_{R} \Big(  \frac16 \tr R^2 \tr_R F_i^2 -\frac23 \tr_R F_i^4\Big)\Big\}\ ,
\end{align}
where $\tr$ denotes the trace in the fundamental representation, $\Tr$ the trace in the adjoint, and $\tr_{R}$ the trace in the representation $R$. The constant is $c=(16\times(2\pi)^{3})^{-1}$. The total anomaly polynomial is
\begin{equation}
    I_{8} = P(\psi_{-}) + P(\chi_{+}) + P(B_{\mu\nu}) + P(\lambda_{-}) + P(\psi_{+})
\end{equation}

In some cases, it is possible to cancel the anomaly via the Green-Schwarz-Sagnotti mechanism \cite{Sagnotti:1992qw}. For this to be possible, the anomaly must factorise into the product of 4-forms as
\begin{equation}\label{eq:anomaly_factorisation}
    \frac{1}{2\pi}I_{8} = \frac{1}{2}\eta_{\alpha\beta}Y^{\alpha}\wedge Y^{\beta}\ ,
\end{equation}
where $\eta_{\alpha\beta} = \diag (+,-,...,-)$ is the invariant tensor of $\SO{1,n_{T}}$ and $Y^{\alpha}$ is a 4-form given by\footnote{Note that the factorisation of the anomaly polynomial is more usually written in terms of $Y^{\alpha} = +\tfrac{1}{2}a^{\alpha}\tr R^{2}+ ...$ However, since we use heavily their results, we will follow the conventions of \cite{Hamada:2023zol} in which they define $a$ with a relative minus sign to simplify some equations later.}
\begin{equation}\label{eq:Y}
    Y^{\alpha} = \frac{1}{(4\pi)^{2}} \left(-\frac{1}{2}a^{\alpha}\tr R^{2} + \sum_{i} b^{\alpha}_{i}\frac{2}{\lambda_{i}}\tr F_{i}^{2} \right) 
    \ .
\end{equation}
The coefficient $\lambda_{i}$ depends on the group and is chosen so that the smallest charge of an embedded $\SU{2}$ instanton is 1. The values for each of the groups are
\begin{align}
&& A_n && B_n && C_n && D_n && E_6 && E_7 && E_8 && F_4 && G_2 
\nonumber\\
\lambda && 1 && 2 && 1 && 2 && 6 && 12 && 60 && 6 && 2
\end{align}
If \eqref{eq:anomaly_factorisation} holds, then one can cancel the anomaly by adding a term to the action of the form
\begin{equation}\label{eq:GSS}
    S_{\text{GSS}} = \eta_{\alpha\beta}B^{\alpha}\wedge Y^{\beta}
\end{equation}
then the gauge transformation of the $B$-field, given by\footnote{Here $b^{\alpha} \in \Omega^{1}$ is a 1-form gauge potential for $B^{\alpha}$, $\Theta$ is local frame rotation, and $\Lambda_{i}$ is a local gauge transformation of the gauge group $G_{i}$.}
\begin{equation}\label{eq:anom_gauge_transf}
    \delta B^{\alpha} = \dd b^{\alpha} - \frac{1}{2}a^{\alpha}\tr (\Theta R) + \sum_{i}b_{i}^{\alpha} \frac{2}{\lambda_{i}}\tr (\Lambda_{i}F_{i})\ ,
\end{equation}
precisely cancels the anomaly obtained from the descent procedure from $I_{8}$.

From \eqref{eq:anomaly_factorisation}, we get clear constraints on the coefficients of $I_{8}$ for the anomaly to be cancellable. Firstly, the $\tr R^{4}$ and $\tr F_{i}^{4}$ terms must vanish. From this, we find
\begin{equation} \label{eq:anom_cancel_1}
    \tr R^{4} \, : \ \ n_{H} - n_{V} + 29n_{T} = 273 \ , \qquad \tr F_{i}^{4}\, : \ \ a^{i}_{\text{adj}} - \sum_{R} n_{R} a^{i}_{R} = 0 \ .
\end{equation}
The first puts constraints on the number of the different multiplets in the theory. The second constrains the representation content. In the second expression, the quantity $n_{R}$ is the number of hypermultiplets in representation $R$ of group $G_{i}$, which can take half-integer values if the representation $R$ is pseudo-real, and $a^{i}_{R}$ is a group theoretical coefficient defined by
\begin{align}
    \tr_{R} F_{i}^{4} &= a^{i}_{R} \tr F_{i}^{4} + b^{i}_{R} \tr F_{i}^{2} \ , \\
    \tr_{R} F_{i}^{2} &= c^{i}_{R} \tr F_{i}^{2} \ .
\end{align}
The remaining constraints require there to be a solution $(a^{\alpha},b_{i}^{\alpha})$ of the equations
\begin{align}
    a\cdot a &= 9 - n_{T}\ , \label{eq:A} \\
    a\cdot b_i &= \frac{1}{6}  \lambda_i \bigg(\sum_{R} n_{R} c^{i}_{R} - c^{i}_{\text{adj}} \bigg)\ , \label{eq:Bi}\\
b_i \cdot b_i &= \frac13  \lambda_i^2 \bigg(\sum_{R} n_{R} b^{i}_{R} - b^{i}_{\text{adj}}\bigg)\ , \label{eq:Cii}\\
b_i \cdot b_j &= \lambda_i\lambda_j \sum_{R,S} n^{i,j}_{R,S}c^{i}_{R} c^{j}_{S}\ . \label{eq:Cij}
\end{align}
In the expressions above, $x\cdot y$ denotes the inner product of $x,y\in \bbR^{n_{T}+1}$ with respect to $\eta$. In the third line there is no summation over $i$, and in the fourth line we take $i\neq j$. These can be solved if and only if the matrix of anomaly coefficients has rank at most $n_{T} + 1$, with at most 1 positive eigenvalue.

It is convenient to write these factors in an $(n+1)\times (n+1)$ matrix $\mathbb{G}$, where $n$ is the number of simple gauge group factors. We will refer to $\mathbb{G}$ as the Gram matrix. Writing $A$ for \eqref{eq:A}, $B_{i}$ for \eqref{eq:Bi}, and $C_{ij}$ for \eqref{eq:Cii} and \eqref{eq:Cij} collectively, we define the Gram matrix as
\begin{equation}\label{eq:Gram}
    \mathbb{G} = \left( \begin{array}{cc}
        A & B_{j} \\
        B_{i} & C_{ij}
    \end{array} \right) \ .
\end{equation}
The absence of local anomalies is then equivalent to the statement that we can write $\mathbb{G}$ as
\begin{equation}\label{eq:local_anom_freedom}
    \mathbb{G} = \mathbb{D}\,\eta\, \mathbb{D}^{T}  \qquad \Leftrightarrow \qquad \mathrm{rank}\,\mathbb{G} \leq n_{T} + 1\,, \ \lambda_{+}(\mathbb{G}) \leq 1\,, \ \lambda_{-}(\mathbb{G}) \leq n_{T}
\end{equation}
where $\mathbb{D}$ is the $(n+1)\times (n_{T}+1)$ matrix of anomaly coefficients $(a^{\alpha},b_{i}^{\alpha})$ and $\lambda_{\pm}$ are the number of positive/negative eigenvalues of $\mathbb{G}$.

Note that it has been shown \cite{Kumar:2010ru} that provided \eqref{eq:anom_cancel_1} is satisfied, the coefficients $B_{i},C_{ij}$ are always integers for any simple group $A_{n\geq 3},\, B_{n\geq 3},\, C_{n\geq 2}, \, D_{n\geq 4},\, E_{n\geq 6},\, F_{4}$. For the groups $\SU{2},\, \SU{3},\, G_{2}$, the same is true if certain global anomalies vanish which we review below.

\subsubsection{Global anomaly freedom}

The conditions in \eqref{eq:local_anom_freedom} guarantee that all local anomalies can be cancelled. These local anomalies correspond to gauge transformations which are continuously connected to the identity. One also needs to consider large gauge transformations which are associated to non-trivial elements of $\pi_{6}(G)$. There are 3 non-trivial cases we need to consider.
\begin{equation}
    \pi_{6}(\SU{2}) = \bbZ_{12} \ , \qquad \pi_{6}(\SU{3}) = \bbZ_{6} \ , \qquad \pi_{6}(G_{2}) = \bbZ_{3} \ .
\end{equation}
As with the global $\SU{2}$ anomaly in 4-dimensions \cite{Witten:1982fp}, whether non-trivial large gauge transformations lead to an anomaly depends on the fermion content of the theory. Indeed, such large gauge transformations contribute a phase to the fermion determinant, and hence the path integral is well-defined only if this phase is equal to 1. To ensure the global anomalies vanish, we must take the following restrictions on the representation content \cite{Bershadsky:1997sb,Suzuki:2005vu}.
\begin{align}
    1-4\sum_{R}n_{R}b_{R} &= 0\mod 3 \qquad G_{2} \label{eq:global_G2} \\
    -2\sum_{R} n_{R}b_{R} &= 0\mod 6 \qquad \SU{3} \label{eq:global_SU3}\\
    4-2\sum_{R}n_{R}b_{R} &= 0 \mod 6 \qquad \SU{2} \label{eq:global_SU2}
\end{align}
In this work we will not consider gauge groups containing $\SU{2}$ or $\SU{3}$ and so we only impose the first condition above. Note that these conditions imply that the Gram matrix $\bbG$ is integral for these gauge groups as well.

A more modern treatment of global anomalies requires also understanding when the Green-Schwaz-Sagnotti anomaly cancelling term is globally well-defined \cite{Monnier:2018cfa,Monnier:2018nfs}. For this to be the case, we require that the cobordism group $\Omega^{\text{spin}}_{7}(BG)$ = 0, where $BG$ is the classifying space of $G$. This seems to provide a contradiction with the conditions \eqref{eq:global_G2} - \eqref{eq:global_SU2} since we have
\begin{equation}\label{eq:cobordism}
    \Omega^{\text{spin}}_{7}(BG_{2}) = \Omega^{\text{spin}}_{7}(B\SU{3}) = \Omega^{\text{spin}}_{7}(B\SU{2}) = 0
\end{equation}
Hence, one might conclude that there are no global anomalies in this case. However, careful consideration of the Green-Schwarz-Sagnotti term shows that \eqref{eq:global_G2} - \eqref{eq:global_SU2} follow from \eqref{eq:cobordism} \cite{Davighi:2020kok}. We will not consider this stricter constraint on global anomaly cancellation, and just impose \eqref{eq:global_G2}.

\subsubsection*{Ghost freedom}

For the supergravity theory to be consistent, there must be some region in the tensor branch in which the theory is ghost free. The scalars in the tensor multiplets parameterise the moduli space $\SO{1,n_{T}}/\SO{n_{T}}$, and can be expressed as some vector $j \in \bbR^{1,n_{T}}$ such that $j^{2} = 1$. Once the Green-Schwarz-Sagnotti anomaly cancelling term \eqref{eq:GSS} is added to the action, supersymmetry fixes the couplings of the kinetic terms for the gauge fields in terms of the anomaly coefficients $b_{i}$, and the tensor multiplet VEV's $j$ (see e.g. \cite{Becker:2023zyb}). More precisely, the coupling for the gauge group $G_{i}$ is proportional to $j\cdot b_{i}$. For the theory to be ghost free, we require that there is some $j\in \bbR^{n_{T}+1}$ such that
\begin{equation}
j^{2} = 1 \ , \qquad j\cdot b_{i} > 0 \ \ \forall\,i \ , \qquad j\cdot a > 0 \ .
\end{equation}
The final term does not imply ghost freedom but instead ensures that the coefficient of the Gauss-Bonet term is positive.

\subsubsection*{Charge lattice consistency}

In 6-dimensional supergravity, there are string states which couple to the antisymmetric $B$ fields in the gravitational and tensor multiplets. Because of the self-duality properties of these fields, the string states will be dyonic, being both electrically and magnetically charged under the $B$-fields. The set of allowed charges must form a lattice which is unimodular, or self-dual, with inner-product of signature $(1,n_{T})$ \cite{Deser:1997se,Seiberg:2011dr}. We will write the charge lattice as $\Lambda_{S} \subset \bbR^{1,n_{T}}$. The self-duality arises by the same arguments as Dirac quantisation of electric/magnetic charges in 4-dimensions.

We can combine this with the completeness hypothesis to get additional constraints on the anomaly coefficients $a,b_{i}$ \cite{Monnier:2017oqd}. We use the generalised completeness hypothesis which states that we should be able to reduce the theory on any spin manifold, and allow any smooth configuration of gauge fields in the path integral. This allows us to consider the integral of the anomaly canceling 4-form $Y^{\alpha}$, given in \eqref{eq:Y}, over any spin 4-manifold $\Sigma_{4}$. Rewriting $Y^{\alpha}$ in terms of characteristic classes, we have
\begin{equation}
    \int_{\Sigma_{4}} Y^{\alpha} = \int_{\Sigma_{4}} \frac{1}{4}a^{\alpha}p_{1} + \sum_{i} b_{i}^{\alpha}c_{2}(F_{i}) = Q^{\alpha}
\end{equation}
where
\begin{equation}
    p_{i} = -\frac{1}{8\pi^{2}}\tr R^{2} \ , \qquad c_{2}(F_{i}) = \frac{1}{8\pi^{2}\lambda_{i}} \tr F_{i}^{2} \ .
\end{equation}
This integral provides a global charge $Q^{\alpha}$ to the theory which, if it is not sourced, generates a global symmetry. This violates the no global symmetries conjecture of quantum gravity and hence $Q^{\alpha}$ must be sourced by a dyonic string wrapping the Poincar\'e dual 2-cycle to $\Sigma_{4}$. That is, we must have $Q^{\alpha}\in \Lambda_{S}$. By demanding this to be true for arbitrary spin manifolds $\Sigma_{4}$, we find that
\begin{equation}\label{eq:characteristic}
    a,b_{i} \in \Lambda_{S} \ , \qquad a\cdot x = x\cdot x\!\! \mod 2 \quad \forall \, x\in \Lambda_{S}\ ,
\end{equation}
where $a\cdot x$ denotes the inner product in the unimodular charge lattice. That is $a$ and $b_{i}$ must be elements of the charge lattice for all $i = 1,...,n$, and $a$ must be a characteristic element.

In practice, once we find the anomaly coefficients $a,b_{i}$ by solving \eqref{eq:anomaly_factorisation}, we check whether the lattice $\Gamma = \mathrm{span}_{\bbZ}(a,b_{i})$ can be embedded into a unimodular lattice of signature $(1,n_{T})$. Such lattices have been classified. For $n_{T} \neq 1\mod 8$, there is a unique (up to $\mathrm{O}(1,n_{T})$ transformations) unimodular lattice with Lorentzian signature, which is odd. We denote it by $\text{I}_{(1,n_{T})}$ and it is given by
\begin{equation}
    \text{I}_{(1,n_{T})} \simeq \bbZ^{1+n_{T}} \ , \qquad \eta = \diag(+,-,...,-).
\end{equation}
For $n_{T}= 1\mod 8$ there are 2 inequivalent lattices. One is the odd lattice $\text{I}_{(1,n_{T})}$ described above. The other is an even lattice, which we denote by $\text{II}_{(1,n_{T})}$, given by
\begin{equation}
    \text{II}_{(1,8k+1)} = U\oplus \underbrace{(-E_{8})\oplus ...\oplus (-E_{8})}_{k} \ , \qquad \eta|_{U} = \left( \begin{array}{cc}
        0 & 1 \\
        1 & 0
    \end{array} \right)
\end{equation}
where $U \simeq \bbZ^{2}$ and $(-E_{8})$ denotes the root lattice of $E_{8}$ with reversed signature.

Whether we can embed $\Gamma = \mathrm{span}_{\bbZ}(a,b_{i})$ into one of these lattices can be determined from the Gram matrix $\bbG$ \cite{VVNikulin_1980}. We will not review these conditions here but instead refer the interested reader to \cite[Appendix B.1]{Hamada:2023zol} for a good review.

\subsubsection*{Probe branes}

The completeness hypothesis states that, in a consistent theory of gravity, every site in the charge lattice should be populated by physical states \cite{Polchinski:2003bq}. In 6-dimensions, this means that there should be string states that one can insert into correlators for every charge $Q\in \Lambda_{S}$. At the level of the low energy supergravity theory, we should be able to insert string defects into the theory which couples to the $B$-fields by adding a term to the action of the form
\begin{equation}
    S_{\text{string}} = \int_{\Sigma} \eta_{\alpha\beta}Q^{\alpha}B^{\beta} + ...
\end{equation}
where $\Sigma$ is the string worldsheet and $...$ denotes terms localised to the worldsheet, the precise details of which are not important. Consistency then requires that the coupled 6d-2d system is anomaly free and unitary \cite{Kim:2019vuc}.

Because of the anomalous gauge transformation of the $B$-fields given by \eqref{eq:anom_gauge_transf}, there is an anomaly inflow onto the string. For the coupled system to be anomaly free, this inflow anomaly should be cancelled by anomalies arising from degrees of freedom localised to the string. Provided the charge $Q$ satisfies some constraints which we describe later, the string is described at low energies by an $\mathcal{N} = (0,4)$ SCFT with global flavour currents associated to the bulk gauge groups $G_{i}$, and a left-moving $\SU{2}$ flavour current \cite{Haghighat:2015ega}. There is also a right-moving $\SU{2}_{R}$ R-symmetry current. The anomaly polynomial $I_{4}$ of the worldsheet theory can be described in terms of the left and right moving central charges $c_{l},c_{r}$, and the levels of the flavour currents $k_{i}, k_{l}$ as
\begin{equation}
    I_{4} = -\frac{c_{r}-c_{l}}{24}p_{1}(T\Sigma) + k_{l}\,c_{2}(F_{l}) - \frac{c_{r}}{6} c_{2}(F_{R}) + k_{i}\, c_{2}(F_{i})
\end{equation}
From this anomaly polynomial, once the contribution from the decoupled center of mass sector have been removed, one can identify the central charges and current levels of the worldsheet theory in terms of the string charge $Q$ \cite{Kim:2019vuc}. These are given by (see \cite{BST} for a review)
\begin{align}
    c_{l} &= 3Q\cdot Q + 9 Q\cdot a + 2 \\
    c_{r} &= 3Q\cdot Q + 3 Q\cdot a \\
    k_{l} &= \tfrac{1}{2}(Q\cdot Q - Q\cdot a + 2) \\
    k_{i} &= Q\cdot b_{i}
\end{align}
Unitarity of the $\mathcal{N} = (0,4)$ SCFT with left-moving flavour currents put constraints on the current levels and central charges of the theory, including that all of the expressions above should be positive. Hence, we restrict the space of $Q$ we consider to be those such that
\begin{equation}\label{eq:inflow_cond_1}
    c_{l} \geq 0 \ , \quad c_{r} \geq 0 \ , \quad k_{l} \geq 0 \ , \quad k_{i} \geq 0
\end{equation}
We also require that the tension of the string is positive. The tension is determined by the charge $Q$ and the VEV of the scalars in the tensor multiplet, parameterised by the vector $j$. The positive tension condition means that we require
\begin{equation}\label{eq:inflow_cond_2}
    T \sim Q\cdot j \geq 0
\end{equation}

Finally, we need to impose that the string is not a string in a Little String Theory (LST), nor does it decompose into a product of instanton strings. In the former case, there is an accidental right-moving $\SU{2}$ symmetry which actually plays the role of R-symmetry, making our identification of $c_{r}$ above incorrect. We therefore exclude this possibility. In the latter case, the worldsheet theory has additional right-moving bosonic modes associated to the moduli space of $G$-instantons which also contribute to $k_{i}$. At generic points on the tensor branch the string will be irreducible, unless the string corresponds to a single instanton string. If this is the case then $Q \sim b_{i}$ and so we restrict $Q$ to be
\begin{equation}\label{eq:inflow_cond_3}
    Q \neq nb_{i} \quad  \forall n\in \bbZ_{> 0}\, , \ \forall i
\end{equation}

Given any probe string in a 6-dimensional supergravity theory with charge $Q$ which satisfies the conditions \eqref{eq:inflow_cond_1}, \eqref{eq:inflow_cond_2}, and \eqref{eq:inflow_cond_3}, for the supergravity theory to be consistent, the coupled 6d-2d theory should be consistent. The worldsheet theory has left-moving flavour currents which will contribute to the left-moving central charge as
\begin{equation}
    c_{i} = \frac{k_{i}\dim G_{i}}{k_{i} + \check{h}_{i}}
\end{equation}
where $\check{h}_{i}$ is the dual coxeter number of the group $G_{i}$. The worldsheet theory is unitary only if the left-moving central charge is large enough to accommodate these contributions. That is, we require\cite{Kim:2019vuc}
\begin{equation}\label{eq:inflow_unitarity}
    c_{l} \geq \sum_{i} c_{i} = \sum_{i} \frac{k_{i}\dim G_{i}}{k_{i} + \check{h}_{i}} \ .
\end{equation}
On the contrary, if we find a $Q$ which satisfies the constraints \eqref{eq:inflow_cond_1} - \eqref{eq:inflow_cond_3} but violates the condition \eqref{eq:inflow_unitarity}, the theory must be inconsistent.

\subsection{Summary of the 6-dimensional supergravity dataset}

In \cite{Hamada:2023zol}, the authors provided an extensive dataset of supergravity models (i.e. gauge groups, representation content,\footnote{Note that in their work and ours, the singlets will always be left implicit. This is because, provided $\dim H^{\text{charged}} - \dim G + 29 n_{T} \leq 273$, one can always add enough singlets to ensure the model is anomaly free.} and a minimum value of $n_{T}$) which can be used as building blocks for anomaly-free 6-dimensional supergravity theories. Their datasets can be found at \cite{GLogesData1,GLogesData2}. In their work, they enumerated over 26 million supergravity models with the following properties.
\begin{itemize}
    \item They allow for any number of tensor multiplets $n_{T}$.

    \item When $n_{T} = 0$ they take any group constructed from $A_{n\geq 3}$, $B_{n\geq3}$, $C_{n\geq2}$, $D_{n\geq 4}$, $E_{n\geq 6}$, $F_{4}$, and $G_{2}$.

    \item When $n_{T}>0$ they take any gauge group constructed from $A_{4\to24}$, $B_{3\to16}$, $C_{3\to16}$, $D_{4\to16}$, $E_{n\geq 6}$, $F_{4}$, and $G_{2}$.

    \item To avoid infinite families, they do not allow certain hypermultiplet content associated to $E_{n}$ gauge group factors. Explicitly, they do not allow $E_{n}$ or $F_4$ factors in the gauge group under which everything is neutral. They also do not allow any $E_{7}$ factors under which only a half-hypermultiplet $\tfrac{1}{2}\textbf{56}$ is charged.

    \item They assume all hypermultiplets are charged under at most 2 simple gauge group factors.

    \item The models are ghost free, and satisfy the charge lattice consistency conditions. When the gauge group contains $G_{2}$, the global anomaly condition \eqref{eq:global_G2} is satisfied.

    \item The models satisfy all conditions for local anomaly freedom, except that the $\tr R^{4}$ anomaly may not vanish. However, any listed model with non-vanishing $\tr R^{4}$ anomaly can be trivially combined (i.e. with no cross-representation content) with other models in the set to form a theory in which anomalies can be cancelled.

    \item They \emph{do not} impose consistency with insertions of probe branes.
\end{itemize}
Their work is complete for the case of $n_{T} = 0$, finding \emph{all} possible models satisfying the above criteria. In the case of $n_{T} = 1$, an extensive study was performed in \cite{Hamada:2024oap} where consistency with probe branes was also analysed.

Their methods to enumerate all such models worked by representing supergravity theories graphically. A similar approach was used for the case $n_{T} = 1$ in \cite{Becker:2023zyb}. The key observation is that one can decompose an anomaly-free model with $k$ simple gauge groups into $k$ theories with only simple gauge groups. For example, the following theory which is anomaly free for $n_{T} \geq 1$.
\begin{equation}\label{eq:example_1}
    \SU{10}\times \SU{16}: \ (\textbf{10},\textbf{16}) + 22(\textbf{1},\textbf{16}) + 2(\textbf{45},\textbf{1}) \ \longrightarrow \ \begin{cases}
        \SU{10}:\ 16\times \textbf{10} + 2\times \textbf{45} \\
        \SU{16}:\ 32\times\textbf{16}
    \end{cases}
\end{equation}
Note that, since the original theory was anomaly free, the individual components in the decomposition must satisfy the quartic casimir condition in \eqref{eq:anom_cancel_1}, and the factorisation condition \eqref{eq:anomaly_factorisation}. They must also be ghost free and satisfy the charge lattice consistency conditions described in the previous section. Note however that we have
\begin{equation}
    \Delta_{i} := \dim H^{\text{charged}}_{i} - \dim(G_{i}) = \begin{cases}
        151 & \SU{10} \\
        257 & \SU{16}
    \end{cases}
\end{equation}
where $H^{\text{charged}}_{i}$ are the hypermultiplets charged non-trivially under $G_{i}$ (so excluding singlets). In each case, we have $\Delta_{i} \geq 273 - 29n_{T}$. Hence, the decomposed models need not satisfy the gravitational anomaly constraint $n_{H} - n_{V} + 29n_{T} = 273$ in \eqref{eq:anom_cancel_1}. Models satisfying all but the gravitational anomaly constraint for some value of $n_{T}$ are referred to as `admissible', and any model which satisfies \emph{all} consistency requirements (anomaly freedom, ghost freedom, charge lattice consistency) as `anomaly free'. Any model with only a simple gauge group will be called simple models, and models with $k$ simple factors will be called $k$-models.

We represent admissible simple models as nodes in a graph. Two nodes are connected by an edge if there is some way to form joint representations such that the resulting 2-model is also admissible. There can be multiple edges connecting each node representing different joint hypermultiplet content, and self-links from a node to itself. Through this method we get a graph where $k$-models are represented by totally connected $k$-sub-graphs, or $k$-cliques.\footnote{It is a quick check to confirm that a $k$-model is admissible if and only if all sub 2-models are admissible.} In principal, the total graph is infinite, since we can have nodes representing exceptional groups with almost arbitrary representation content. To obtain a finite graph, \cite{Hamada:2023zol} imposed bounds on the maximum value of $\Delta(G_{i})$ for each simple gauge group $G_{i}$. Once this is done, clique finding algorithms were used to find \emph{all} admissible $k$-models in the graph. 

The list of models given in the datasets \cite{GLogesData1,GLogesData2} are the irreducible admissible models. Irreducible means that all nodes in the corresponding $k$-clique are connected through edges which correspond to non-trivial joint hypermultiplets. That is, the model does not correspond to the trivial sum of two smaller models. For example, the model \eqref{eq:example_1} is irreducible, but the model
\begin{equation}
    \SU{10}\times \SU{10}:\ 20(\textbf{10},\textbf{1}) + 20(\textbf{1},\textbf{10})
\end{equation}
is not irreducible since it is the trivial product of two copies of $\SU{10}:\ 20\times \textbf{10}$. Moreover, the models provided satisfy the further condition that, if they are not anomaly free, they can be combined trivially with other models in the set such that the resulting theory is anomaly free.

The data given in \cite{GLogesData1,GLogesData2} consists of 26,760,256 entries describing the irreducible admissible models found. The data listed about each model are
\begin{itemize}
    \item The simple gauge group factors.

    \item The (joint) representation content of the model.

    \item The Gram matrix entries.

    \item The minimum value $T_{\text{min}}$ of $n_{T}$ for which the model is anomaly free.

    \item The value of $\Delta := \dim H^{\text{charged}} - n_{V}$ for the total model.

    \item The number of positive and negative eigenvalues for the submatrix $C_{ij}$ of the Gram matrix as in \eqref{eq:Gram}.
\end{itemize}

\section{Machine Learning and Autoencoders}\label{ML-Auto}

Machine learning has gained traction in studies of the string landscape in recent years. One of the reasons for this is that solutions to string theory naturally form enormous datasets which can be difficult for humans to make sense of unaided. Machine learning, and in particular unsupervised learning algorithms, can be used to great effect as they can find patterns in large sets of data with minimal human input. In this section, we will apply a particular unsupervised learning algorithm called an autoencoder, which is reviewed in e.g.~\cite{Hinton2006} and (specifically for a physics audience) in \cite{Ruehle:2020jrk}, to study the large dataset of anomaly-free supergravities in 6-dimensions. The aim is to better understand the string landscape not from a top-down approach, which has been the focus of many previous works, but from a bottom-up swampland perspective in 6-dimensions.

Briefly, an autoencoder is a fully-connected feed-forward network in which the input and output layers are identical, and the hidden layers separating them consist of fewer dense nodes than those on the ends.  The data of interest is used as both input and output, so the goal is for the data to pass through the network as unchanged as possible.  While the most straightforward way for this to happen would be for the network to learn the identity function, the smaller layers in the middle create an information bottleneck which makes it impossible for the network to fully learn the identity function -- instead, it must get as close as it can.  The result, then, is that the data must be compressed in the middle layers, and the optimization procedure will force this compression to become more efficient.  In practice, given a large set of data, this process leads to an algorithm which dynamically selects for similarity between the data points, effectively creating groups within which it sorts the data, without these categories needing to be predetermined by the user.  This auto-classification is thus a form of unsupervised machine learning. Such a process has been used many times before for organizing string theory data.  See \cite{Mutter:2018sra,Halverson:2019tkf,Deen:2020dlf,escalantenotario2024autoencoderheteroticorbifoldsarbitrary,Otsuka_2020,ishiguro2023autoencoderdrivenclusteringintersectingdbrane} for a few examples, of which we will follow \cite{Mutter:2018sra} most closely. Another prominent use of autoencoders is `anomaly detection'. Each pass of data through the autoencoder comes with a reconstruction loss as the NN cannot perfectly reconstruct the input data from the compressed layers. If an input is very similar to most other inputs, then its reconstruction loss should be close to the average. In contrast, if some data is very different to the majority of the inputs, i.e.~represents an anomaly in the data, then it should have a relatively large reconstruction loss. In this way, the autoencoder can identify anomalies without human input as well. Here, anomaly refers to an input which is unlike the others, or peculiar in some way. So to not confuse with the anomaly associated to quantum theories discussed in section \ref{sec:consistency_review}, we shall instead refer to this as `peculiarity detection', and the data that the autoencoder identifies as `peculiar'.

In this section, we use both applications of autoencoders to study 6-dimensional supergravities. We shall discuss the precise architecture we chose and training of the model, before discussing the classification we obtain through clustering analysis. We then use the autoencoder to identify a handful of interesting models through the peculiarity detection and analyse their properties in more detail.

\subsection{Autoencoder Architecture and Training}\label{sec:ae_training}

Autoencoders are neural nets which consist of two parts -- an encoder and a decoder. The encoder takes some input and transforms it into some other form, the output of which we call the latent layer. The decoder then takes this transformed data and tries to reconstruct the input as accurately as possible. Both the encoder and decoder are trained simultaneously, where the total output is the output of encoder and decoder combined, and the loss function measures the similarity between the input and output. In our application, we took the loss function to be the mean squared logarithmic error (MSLE). That is, if the input $\mathbf{x}$ and output $\hat{\mathbf{x}}$ of the autoencoder are of dimension $k$, then the loss function is\footnote{Typically, during training the data is fed through the autoencoder in batches. The loss function is then averaged over the batch.}
\begin{equation}
    L(\textbf{x},\hat{\textbf{x}}) = \frac{1}{k}\sum_{i=1}^{k} (\log(1+x_{i}) - \log(1+\hat{x}_{i}))^{2}
\end{equation}
By tuning the parameters in the autoencoder, one can minimise the loss function. In doing so, one ensures that the latent layer is as true a representation of the input data as possible.

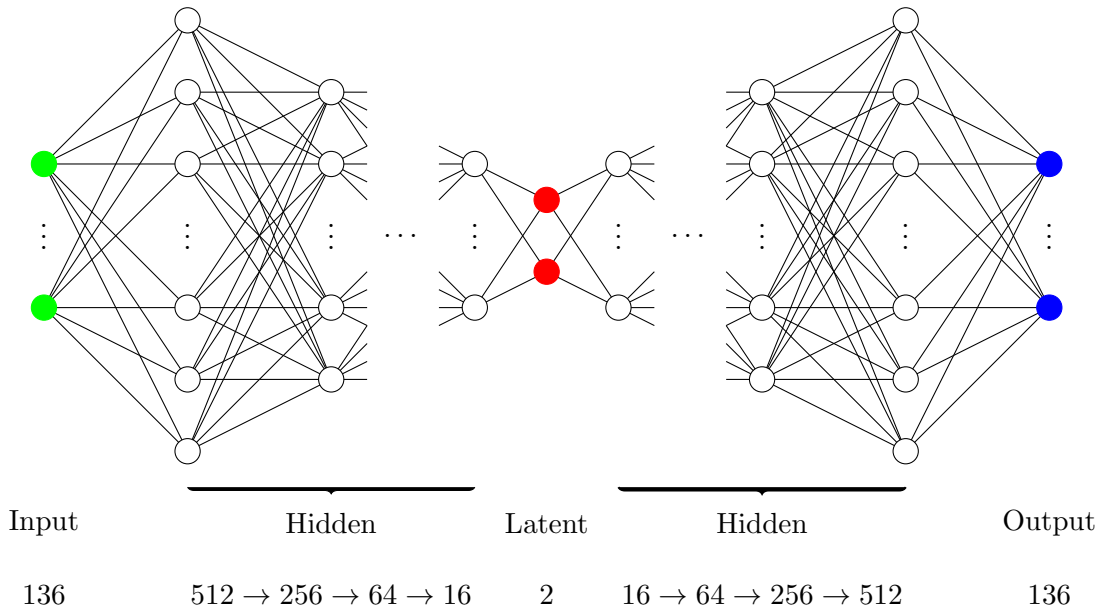
\begin{figure}[h]
    \centering
\begin{tikzpicture}[roundnode/.style={circle, thick, minimum size=7mm}, scale=0.95]        
        %
        %
        \draw (-7,1) -- (-5,3);
        \draw (-7,1) -- (-5,2);
        \draw (-7,1) -- (-5,1);
        \draw (-7,1) -- (-5,-1);
        \draw (-7,1) -- (-5,-2);
        \draw (-7,1) -- (-5,-3);
        \draw (-7,-1) -- (-5,3);
        \draw (-7,-1) -- (-5,2);
        \draw (-7,-1) -- (-5,1);
        \draw (-7,-1) -- (-5,-1);
        \draw (-7,-1) -- (-5,-2);
        \draw (-7,-1) -- (-5,-3);
        %
        %
        %
        \filldraw[color=green, fill=green] (-7,1) circle (5pt);
        \node[roundnode] at (-7,0.1) {$\vdots$};
        \filldraw[color=green, fill=green] (-7,-1) circle (5pt);
        %
        \node[roundnode] at (-7,-4) {Input};
        \node[roundnode] at (-7,-5) {136};
        \draw (-5,3) -- (-3,2);
        \draw (-5,3) -- (-3,1);
        \draw (-5,3) -- (-3,-1);
        \draw (-5,3) -- (-3,-2);
        \draw (-5,2) -- (-3,2);
        \draw (-5,2) -- (-3,1);
        \draw (-5,2) -- (-3,-1);
        \draw (-5,2) -- (-3,-2);
        \draw (-5,1) -- (-3,2);
        \draw (-5,1) -- (-3,1);
        \draw (-5,1) -- (-3,-1);
        \draw (-5,1) -- (-3,-2);
        \draw (-5,-1) -- (-3,2);
        \draw (-5,-1) -- (-3,1);
        \draw (-5,-1) -- (-3,-1);
        \draw (-5,-1) -- (-3,-2);
        \draw (-5,-2) -- (-3,2);
        \draw (-5,-2) -- (-3,1);
        \draw (-5,-2) -- (-3,-1);
        \draw (-5,-2) -- (-3,-2);
        \draw (-5,-3) -- (-3,2);
        \draw (-5,-3) -- (-3,1);
        \draw (-5,-3) -- (-3,-1);
        \draw (-5,-3) -- (-3,-2);
        \filldraw[color=black, fill=white] (-5,3) circle (5pt);
        \filldraw[color=black, fill=white] (-5,2) circle (5pt);
        \filldraw[color=black, fill=white] (-5,1) circle (5pt);
        \node[roundnode] at (-5,0.1) {$\vdots$};
        \filldraw[color=black, fill=white] (-5,-1) circle (5pt);
        \filldraw[color=black, fill=white] (-5,-2) circle (5pt);
        \filldraw[color=black, fill=white] (-5,-3) circle (5pt);
        %
        %
        \draw (-3,2) -- (-2.5,2);
        \draw (-3,2) -- (-2.5,1.75);
        \draw (-3,2) -- (-2.5,1.5);
        \draw (-3,2) -- (-2.5,1.25);
        \draw (-3,1) -- (-2.5,1.25);
        \draw (-3,1) -- (-2.5,1);
        \draw (-3,1) -- (-2.5,0.75);
        \draw (-3,1) -- (-2.5,0.5);
        \draw (-3,-1) -- (-2.5,-0.5);
        \draw (-3,-1) -- (-2.5,-0.75);
        \draw (-3,-1) -- (-2.5,-1);
        \draw (-3,-1) -- (-2.5,-1.25);
        \draw (-3,-2) -- (-2.5,-1.25);
        \draw (-3,-2) -- (-2.5,-1.5);
        \draw (-3,-2) -- (-2.5,-1.75);
        \draw (-3,-2) -- (-2.5,-2);
        \node[roundnode] at (-2,0) {$\dots$};
        \filldraw[color=black, fill=white] (-3,2) circle (5pt);
        \filldraw[color=black, fill=white] (-3,1) circle (5pt);
        \node[roundnode] at (-3,0.1) {$\vdots$};
        \filldraw[color=black, fill=white] (-3,-1) circle (5pt);
        \filldraw[color=black, fill=white] (-3,-2) circle (5pt);
        \draw (-1.5,1.25) -- (-1,1);
        \draw (-1.5,1) -- (-1,1);
        \draw (-1.5,0.75) -- (-1,1);
        \draw (-1.5,0.5) -- (-1,1);
        \draw (-1.5,-1.25) -- (-1,-1);
        \draw (-1.5,-1) -- (-1,-1);
        \draw (-1.5,-0.75) -- (-1,-1);
        \draw (-1.5,-0.5) -- (-1,-1);
        \draw (-1,1) -- (0,0.5);
        \draw (-1,1) -- (0,-0.5);
        \draw (-1,-1) -- (0,0.5);
        \draw (-1,-1) -- (0,-0.5);
        \filldraw[color=black, fill=white] (-1,1) circle (5pt);
        \node[roundnode] at (-1,0.1) {$\vdots$};
        \filldraw[color=black, fill=white] (-1,-1) circle (5pt);
        \filldraw[decoration = {brace, mirror}, decorate] (-5,-3.5) -- (-1,-3.5);
        \node[roundnode] at (-3,-4) {Hidden};
        \node[roundnode] at (-3,-5) {$512\to256\to64\to16$};
        %
        \draw (0,0.5) -- (1,1);
        \draw (0,0.5) -- (1,-1);
        \draw (0,-0.5) -- (1,1);
        \draw (0,-0.5) -- (1,-1);
        \filldraw[color=red, fill=red] (0,0.5) circle (5pt);
        \filldraw[color=red, fill=red] (0,-0.5) circle (5pt);
        \node[roundnode] at (0,-4) {Latent};
        \node[roundnode] at (0,-5) {2};
        %
        \draw (1.5,1.25) -- (1,1);
        \draw (1.5,1) -- (1,1);
        \draw (1.5,0.75) -- (1,1);
        \draw (1.5,0.5) -- (1,1);
        \draw (1.5,-1.25) -- (1,-1);
        \draw (1.5,-1) -- (1,-1);
        \draw (1.5,-0.75) -- (1,-1);
        \draw (1.5,-0.5) -- (1,-1);
        \filldraw[color=black, fill=white] (1,1) circle (5pt);
        \node[roundnode] at (1,0.1) {$\vdots$};
        \filldraw[color=black, fill=white] (1,-1) circle (5pt);
        \node[roundnode] at (2,0) {$\dots$};
        \draw (3,2) -- (2.5,2);
        \draw (3,2) -- (2.5,1.75);
        \draw (3,2) -- (2.5,1.5);
        \draw (3,2) -- (2.5,1.25);
        \draw (3,1) -- (2.5,1.25);
        \draw (3,1) -- (2.5,1);
        \draw (3,1) -- (2.5,0.75);
        \draw (3,1) -- (2.5,0.5);
        \draw (3,-1) -- (2.5,-0.5);
        \draw (3,-1) -- (2.5,-0.75);
        \draw (3,-1) -- (2.5,-1);
        \draw (3,-1) -- (2.5,-1.25);
        \draw (3,-2) -- (2.5,-1.25);
        \draw (3,-2) -- (2.5,-1.5);
        \draw (3,-2) -- (2.5,-1.75);
        \draw (3,-2) -- (2.5,-2);
        \draw (5,3) -- (3,2);
        \draw (5,3) -- (3,1);
        \draw (5,3) -- (3,-1);
        \draw (5,3) -- (3,-2);
        \draw (5,2) -- (3,2);
        \draw (5,2) -- (3,1);
        \draw (5,2) -- (3,-1);
        \draw (5,2) -- (3,-2);
        \draw (5,1) -- (3,2);
        \draw (5,1) -- (3,1);
        \draw (5,1) -- (3,-1);
        \draw (5,1) -- (3,-2);
        \draw (5,-1) -- (3,2);
        \draw (5,-1) -- (3,1);
        \draw (5,-1) -- (3,-1);
        \draw (5,-1) -- (3,-2);
        \draw (5,-2) -- (3,2);
        \draw (5,-2) -- (3,1);
        \draw (5,-2) -- (3,-1);
        \draw (5,-2) -- (3,-2);
        \draw (5,-3) -- (3,2);
        \draw (5,-3) -- (3,1);
        \draw (5,-3) -- (3,-1);
        \draw (5,-3) -- (3,-2);
        \filldraw[color=black, fill=white] (3,2) circle (5pt);
        \filldraw[color=black, fill=white] (3,1) circle (5pt);
        \node[roundnode] at (3,0.1) {$\vdots$};
        \filldraw[color=black, fill=white] (3,-1) circle (5pt);
        \filldraw[color=black, fill=white] (3,-2) circle (5pt);
        \filldraw[thick, decoration = {brace}, decorate] (5,-3.5) -- (1,-3.5);
        \node[roundnode] at (3,-4) {Hidden};
        \node[roundnode] at (3,-5) {$16\to64\to256\to512$};
        \draw (7,1) -- (5,3);
        \draw (7,1) -- (5,2);
        \draw (7,1) -- (5,1);
        \draw (7,1) -- (5,-1);
        \draw (7,1) -- (5,-2);
        \draw (7,1) -- (5,-3);
        \draw (7,-1) -- (5,3);
        \draw (7,-1) -- (5,2);
        \draw (7,-1) -- (5,1);
        \draw (7,-1) -- (5,-1);
        \draw (7,-1) -- (5,-2);
        \draw (7,-1) -- (5,-3);
        \filldraw[color=black, fill=white] (5,3) circle (5pt);
        \filldraw[color=black, fill=white] (5,2) circle (5pt);
        \filldraw[color=black, fill=white] (5,1) circle (5pt);
        \node[roundnode] at (5,0.1) {$\vdots$};
        \filldraw[color=black, fill=white] (5,-1) circle (5pt);
        \filldraw[color=black, fill=white] (5,-2) circle (5pt);
        \filldraw[color=black, fill=white] (5,-3) circle (5pt);
        \node[roundnode] at (7,-4) {Output};
        \node[roundnode] at (7,-5) {136};
        \filldraw[color=blue, fill=blue] (7,1) circle (5pt);
        \node[roundnode] at (7,0.1) {$\vdots$};
        \filldraw[color=blue, fill=blue] (7,-1) circle (5pt);
    \end{tikzpicture}
    
    \caption{A diagram of the autoencoder used to study 6-dimensional supergravities. The input was the Gram matrix entries. Since this is a symmetric matrix with at most $16\times 16$ size, the input (\textcolor{green}{green}) and output (\textcolor{blue}{blue}) were both 136 dimensional vectors. The latent layer (\textcolor{red}{red}) was 2-dimensional. There were 4 hidden layers in both the encoder (\textcolor{green}{green}$\to$\textcolor{red}{red}) and decoder (\textcolor{red}{red}$\to$\textcolor{blue}{blue}), with dimensions (512, 256, 64, 16) and (16, 64, 256, 512) respectively. The activation in the hidden layers was ReLU and the loss function was MSLE.}
    \label{fig:autoencoder}
\end{figure}

The precise form of the latent layer and the architecture of the encoder/decoder depends heavily on the application. The state of the art are variational autoencoders where the latent layer is a probability distribution representing the population of data. The decoder samples the distribution to produce an element of the input data. While these can be useful for e.g. generative NN creating images which mimic the natural variability of human creativity, it is not particularly useful for our purposes. In our case, we want the autoencoder to learn the salient features of a fixed and well-defined, albeit complicated, mathematical object. We therefore decided to build a simple, totally connected, feed-forward autoencoder in which the latent layer is a vector of lower dimension than the input. The encoder then compresses the input data forcing the encoder to identify and represent the most salient features. These can be identified through clustering in the latent space. We chose the non-linear activation function in all of the hidden layers of the autoencoder to be the ReLU function. A representative diagram of the final autoencoder is given in Figure \ref{fig:autoencoder}.

There were many important choices when designing the autoencoder. The first was how best to represent our data, i.e. the shape of the input layer to the NN. One could envision many different ways to represent the models. For example, one could use the graphical representation used by the authors in \cite{Hamada:2023zol} to represent each model, denoting the graphs by long vectors of 0's and 1's, the 1's denoting which nodes are present in the graph. Examining the data, we found that there were 14,423 distinct nodes present. This representation of the data would require an input layer of at least that dimension, creating a NN with many 100,000's of parameters. Even with the large amount of data we had, such a large NN, combined with the extremely sparse nature of the input data, would be difficult to train.

Instead, we opted to focus on the information contained in the Gram matrix. This had many benefits. For a start, the Gram matrices contained in the dataset were at most $16\times 16$ symmetric matrices, leading to 136 unique parameters. This meant that we could represent models with 136-dimensional input layer. The Gram matrix also contains, indirectly, information about the groups and representation content of the model, given that the parameters are determined by group theory coefficients (see \eqref{eq:A}-\eqref{eq:Cij}). Finally, many physical properties of the model, including anomaly inflow, are determined from the Gram matrix. Using this as the input to the autoencoder would increase the chances that the NN would learning physically relevant information about the models.

One possible downside to this approach is that different models can have the same Gram matrix and so would not be distinguished by the autoencoder. We expect, however, that these different models would have very similar properties and so not too much is lost by this simplification. One could consider ways to distinguish between these models by, e.g. providing more explicit information about the gauge groups in the input data, but we leave this refinement for future work.

Another important decision was deciding the size of the latent layer. The lower the dimension of the latent layer, the more the autoencoder is forced to learn large-scale features of the space of models. However, this also increases the likelihood that the NN is irrevocably removing important information as it compresses the data. Nonetheless, it seems likely that the form of the input data should allow for fairly large compression without too much corruption of the data. This is because the Gram matrix entries are not 136 independent real variables. Instead, they must be quantised in integer units\footnote{When it comes to training, all inputs were normalised to be between $1$ and $-1$.} and the inputs are not independent. Indeed the $(i,i)$ and $(0,i)$ components of the Gram matrix are determined from the same group and representation content. This would imply that there is a reasonable amount of redundancy in this presentation of the data which would be removed by optimising the autoencoder.

\begin{figure}[h]
    \centering
    \includegraphics[width=0.75\linewidth]{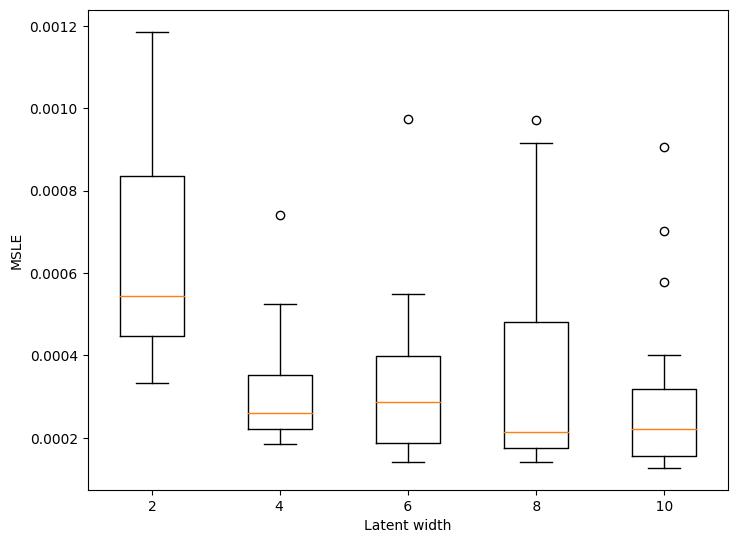}
    \caption{Boxplots for the MSLE of 5 different autoencoders with different widths for the latent layers. The boxplots represent the losses obtained from 20 training runs, each for 20 epochs, for each of the autoencoders.}
    \label{fig:autoenc_loss_boxplot}
\end{figure}

To ensure that we were not over-compressing the data, we ran a series of tests on a subset of the data with different dimensions for the latent layer. We trained on a random sample of around 500,000 data points and trained 5 different autoencoders over 20 epochs, 20 times each. The widths for the latent layer of the 5 different autoencoders were 2, 4, 6, 8, and 10. There were 2 hidden layers in the encoder and decoder with dimension (50, 10) and (10, 50) respectively. A box plot for the losses obatined across the 20 training runs for each of the autoencoders is given in Figure \ref{fig:autoenc_loss_boxplot}.

Over-compression of the data and information loss would be indicated by a significantly larger value of the MSLE for smaller widths of the latent layer. Note from Figure \ref{fig:autoenc_loss_boxplot} that there does appear to be a larger loss for a latent width of 2 compared to the others. This is partially to be expected; the smaller the latent layer the greater the loss tends to be. However, there is no significant difference in loss between latent widths of 4, 6, 8, and 10. We should emphasise, however, that the absolute difference between the MSLE for all of the different autoencoders is very small, approximately $\mathcal{O}(10^{-4})$. Taking the median value for the autoencoder of latent width 2, we get an MSLE of approximately $5.5\times 10^{-4}$ which corresponds to roughly 2.4\% average relative error between the input and the output. Doing the same for the autoencoder with latent width 10, we get a median MSLE of approximately $2.5\times 10^{-4}$ which corresponds to roughly 1.6\% average relative error between input and output. Ultimately, we decided that the benefit of being able to plot the output of the latent layer in a 2-dimensional plot was more important than reducing the error 0.8\%. The final architecture we chose, which gave the smallest average loss while also training in a reasonable amount of time, is given in figure \ref{fig:autoencoder}.

\subsection{Clustering Analysis}\label{sec:analysis}

\begin{figure}
    \centering
    \includegraphics[width=0.85\linewidth]{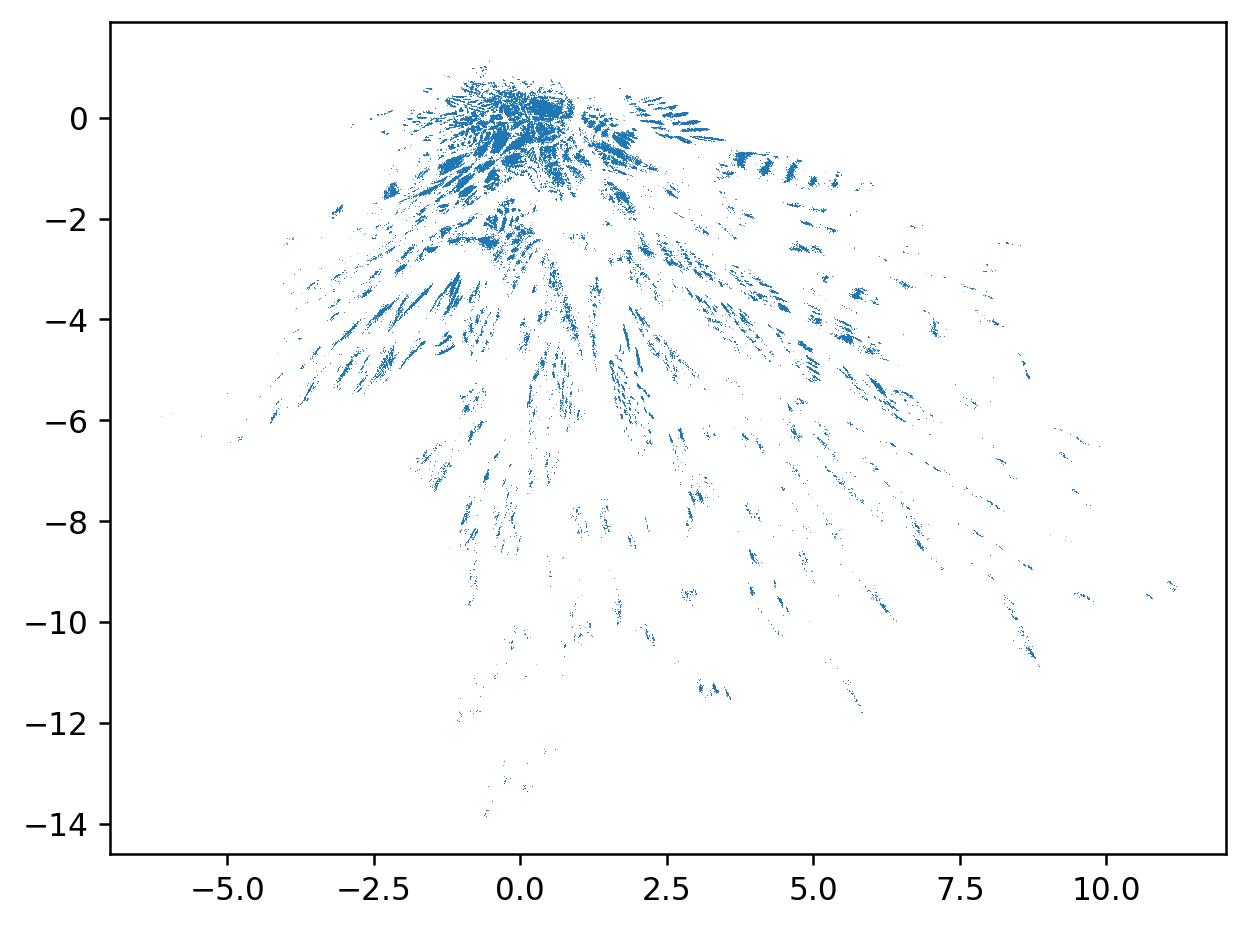}
    \caption{The output of the latent layer of the fully trained autoencoder. Each point in the graph represents a single model or Gram matrix.}
    \label{fig:plain_latent_outp}
\end{figure}

After training the final autoencoder on all of the data, we plotted the output of the 2-dimensional latent layer for all inputs. The result is shown in Figure \ref{fig:plain_latent_outp}.

Note that one should not associate any physical meaning with the precise $x$ and $y$-values in the graph. These are simply labels that the autoencoder has assigned to each model, from which it can reconstruct the original input. What is significant, however, is the presence of clustering. It is clear from Figure \ref{fig:plain_latent_outp} that there is clustering of points, i.e. models, within the latent layer. This implies that the autoencoder may have learned certain features of the space of Gram matrices and grouped models with similar properties together. We cannot understand completely which features the autoencoder has learned, but by studying the properties of these clusters, we can obtain a classification of 6-dimensional supergravity theories that is free from human input. The results of our clustering analysis are provided in \cite{ml-github-repo,ml-github-website}.

\begin{figure}[h!]
\centering
    \begin{subfigure}{0.7\textwidth}
    \includegraphics[width=\linewidth]{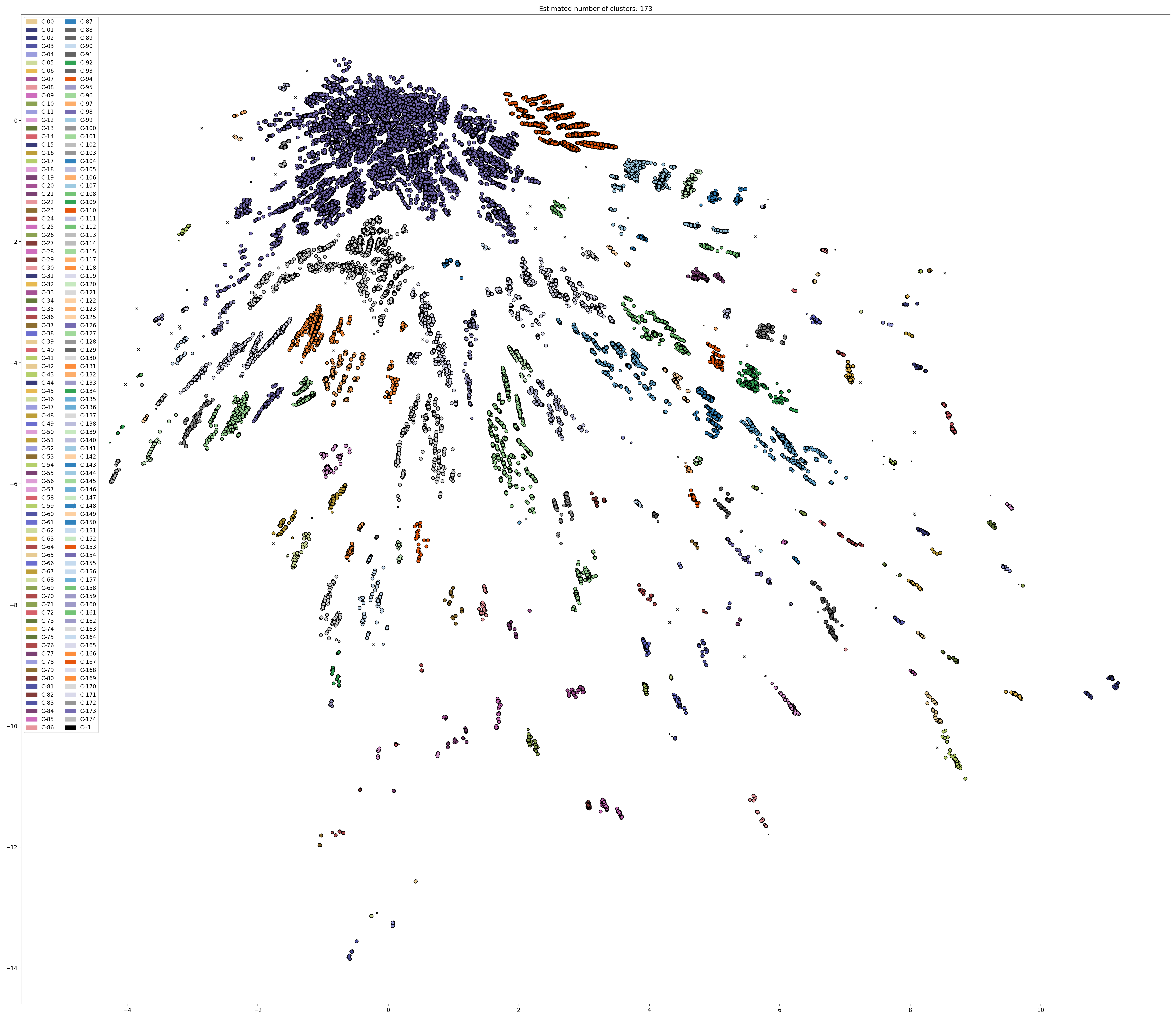}
    \caption{Initial clustering of all models (175 clusters).}
    \label{fig:color_coded_latent}
    \end{subfigure}
    \begin{subfigure}{0.7\textwidth}
    \includegraphics[width=\linewidth]{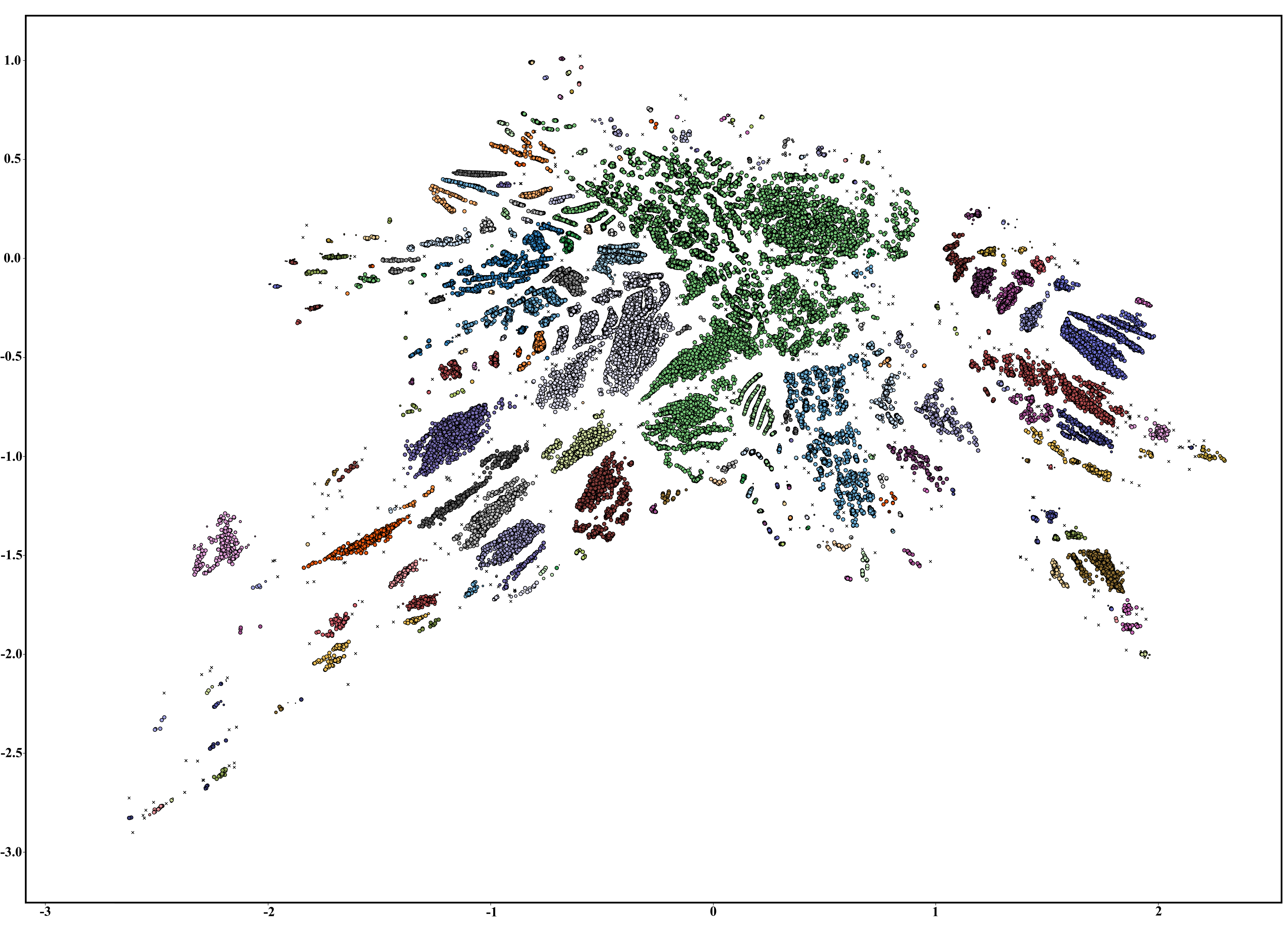}
    \caption{Sub-clustering of C-174 (largest cluster) into a set of 234 sub-clusters.}
    \label{fig:LC-01_clustering}
    \end{subfigure}
    \caption{The output of the latent layer of the fully trained autoencoder clustered by the \texttt{fast-hdbscan} algorithm. The different colors represent the different clusters identified by the algorithm. A description of the clusters is given in \cite{ml-github-website}}
    \label{fig:clustering_latent}
\end{figure}

In order to identify the clusters, we applied the clustering algorithm, \texttt{fast-hdbscan},\footnote{this is a multi-core implementation of scikit-learn's clustering algorithm, \texttt{hdbscan}.} which identifies clusters based on an underlying hierarchy of densities determined by the fully connected minimum spanning tree, where edges of the graph are weighted according to some mutual reachability distance measure relevant to their associated vertices. Clusters are then systematically identified by `cutting' edges with the highest weight and then labeled. The details of the clustering algorithm are not relevant for our work but a more complete description of the algorithm and its development can be found here \cite{scikit:hdbscan}.

In the initial application of the clustering algorithm, a total of 175 clusters were identified (C-1 to C-175).\footnote{Labels have been shifted up by one relative to what's given in the legend. Outliers are labeled with a -1.} A color-coded diagram of the identified clusters is given in Figure \ref{fig:color_coded_latent}, although we provide a more detailed image of the clustered latent layer in \cite{ml-github-website}. The output of the clustering algorithm identified one particularly large cluster, C-174 in the top left-hand corner of Figure \ref{fig:color_coded_latent}, which contained over 14 million points. This cluster comprises over half of the data, so to get a more refined description of the latent space clustering we chose to run the clustering algorithm again on the subset of data in C-174. This identified an additional 100 clusters\footnote{This is after consolidating clusters that were too small to be useful.}, for a total of 275 clusters. The results of the second clustering algorithm are presented in Figure \ref{fig:LC-01_clustering}.

\begin{figure}[h]
    \centering
    \begin{subfigure}{0.7\textwidth}
        \includegraphics[width=\linewidth]{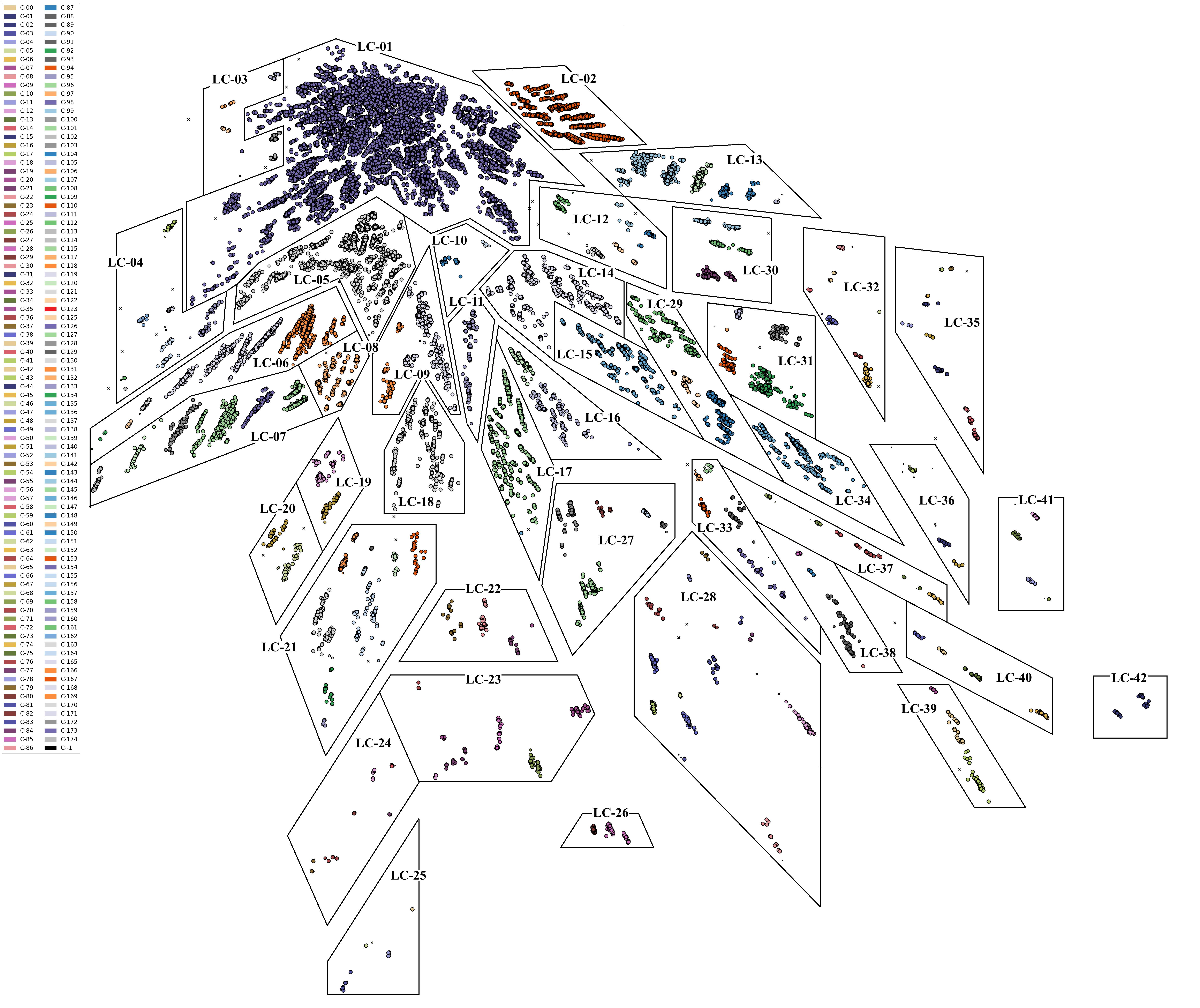}
        \caption{Collecting the latent layer in `local cluster groups' (LC), identified by the black boxes and labelled.}
        \label{fig:local_cluster_groups}
    \end{subfigure}
    \begin{subfigure}{0.7\textwidth}
        \includegraphics[width=\linewidth]{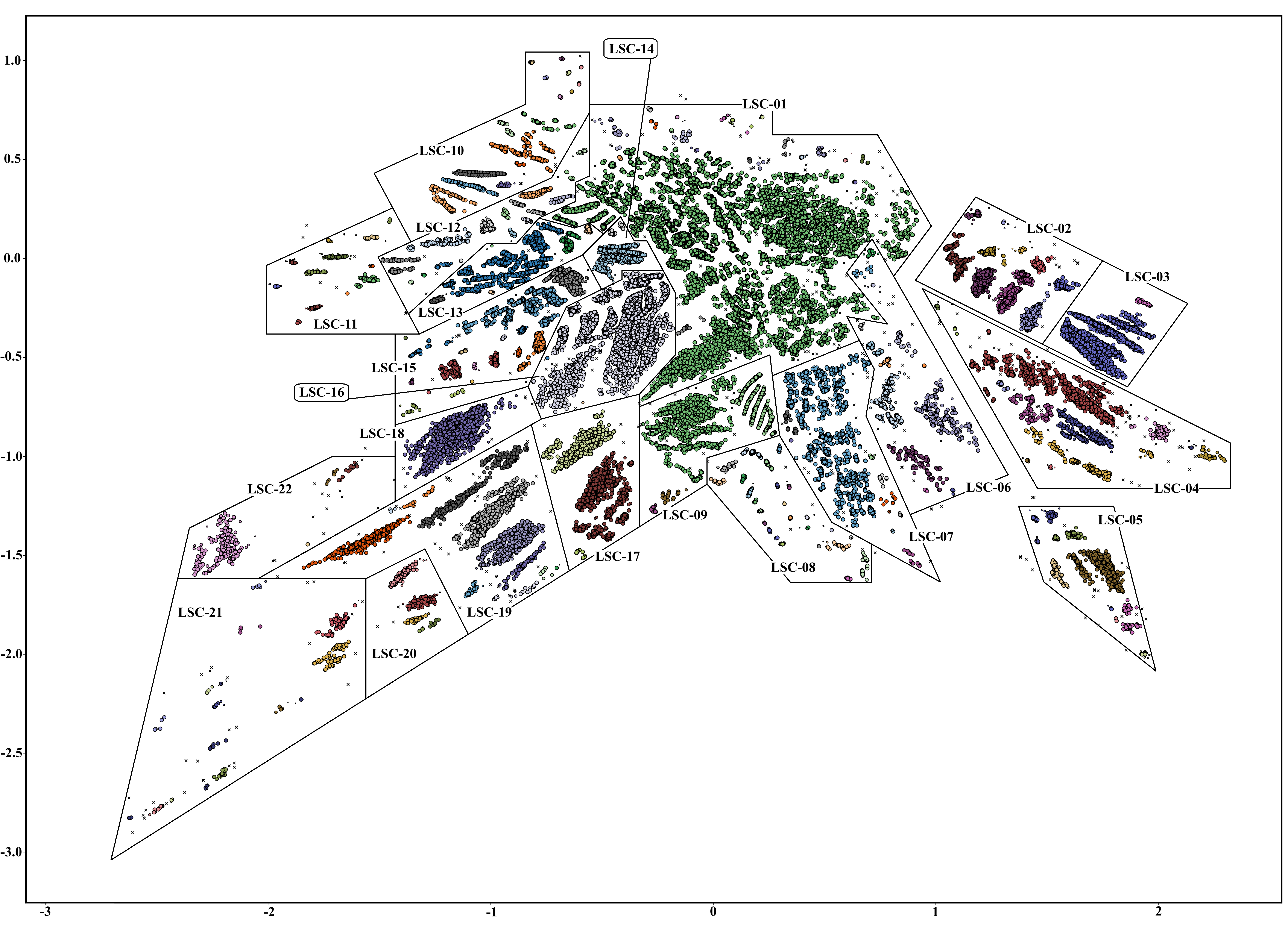}
        \caption{Further partitioning of LC-01 into `Local Sub-Cluster groups' (LSC), identified by the black boxes and labeled.}
        \label{fig:local_sub_cluster_groups}
    \end{subfigure}
\end{figure}

Rather than give an in-depth description of the properties of all identified clusters here, we refer the reader to \cite{ml-github-website,ml-github-repo} where we provide the full composition of features. To aid our presentation there and here we decided to collect the clusters identified by the clustering algorithm into larger `local cluster groups' by hand. We identified 42 local cluster groups (LC-01 - LC-42), and these are shown in Figure \ref{fig:local_cluster_groups} by a black outline and labeled. Note that LC-01 is made entirely of the cluster C-174 which was further split into 'local sub-clusters' by hand in our second application of \texttt{fast-hdbscan} as described above. An additional 22 local sub-cluster groups (LSC-01 - LSC-22) were identified and are shown in Figure \ref{fig:local_sub_cluster_groups} labeled and outlined in black. Generic properties of clusters such as the number of models, types of k-Cliques, most abundant group sub-type, and which Lie-types are present can be found in Table \ref{tab:CGClusters} for each local cluster and similarly in Table \ref{tab:LC1Clusters} for each local sub-cluster. The remaining analysis focuses on highlighting some of the more interesting clusters we found, however, much remains to be interpreted for a more complete analysis. We leave this for the future.

\subsubsection{General analysis}
Our reductive approach to analysing clusters can broadly be divided into three levels of analysis. Each level narrows the focus of the analysis down to subsets of models which share more overlapping features. At each level we analyse the abundance of group factors, their associated gram matrix entries, and identify the graphs associated to the models appearing in these clusters.\footnote{Due to the overwhelming number of available models any attempt to provide the full set of information for every single model would border on cruel and unusual punishment to the reader. Instead we provide the full set of information, including representation content, of each model within a given cluster as a separate downloadable Mathematica file found here \cite{ml-github-repo}.}

At the top level is where (sub-)clusters are grouped into larger Local (Sub-)Clusters. While it's expected that clusters which are relatively close in the latent layer will have similar characteristic features, it's likely that some individual clusters may share more features with another adjacent cluster in a different local group. Level two focuses on clusters identified by \texttt{hdbscan} and is where we find some of the more prominent features begin to surface. Since the data is comprised of models of various sizes ({\it up to 15 gauge group factors}) clusters often include a mixture of different sized models. The last level is where we now separate models according to the number of group factors and one can begin to look at more direct comparisons of gram matrices across the clusters.\footnote{We provide the functionality to more easily compare models across clusters in our website \cite{ml-github-website} using the Comparison View tool. Again, due to the overwhelming variety of models available we highly encourage the reader to explore the clustering data using this tool rather than attempting to open each page separately.}

As was detailed throughout Sec.~\ref{sec:gsugra}, there are a number of features from which the set of admissible/anomaly-free models are comprised. These include: the composition of gauge group factors, their representation content, the minimum number of tensor multiplets, number of vector and hypermultiplets, and the Gram matrix. Since much of these features are encoded within the Gram matrix itself this motivates our decision to use it to train our NNs which we described in the previous subsection \ref{sec:ae_training}. The clustering of models in the 2d latent layer then indicates a high degree of similarity between their respective gram matrices. What consequences this may have for revealing potential patterns within the other model dependent features is left for future work.

\bigskip

\noindent{\it Some generic observations regarding all (sub-)clusters:}
\begin{itemize}
    \item The $A$-type groups appear at least once in every (sub-)cluster

    \item The $B$ and $D$-type groups appear at least once in almost all (sub-)clusters. The exceptions are C-126, SC-57, and SC-71 where the $D$-type groups do not appear, and SC-26, SC-63, and SC-84 where both the $B$ and $D$-type groups do not appear. More details on the (sub-)clusters are below and can also be found in \cite{ml-github-website}.
    
    \begin{center}
    \begin{tabular}{c|c|c}
        \hline
        (Sub-)Cluster & \# of Models & Group Types \\\hline
        C-126 {\it (LC-03)} & 628 & \(\{A,B,C,G\}\)\\
        SC-26 {\it (LSC-05)} & 13,470 & \(\{A,C,G\}\)\\
        SC-57 {\it (LSC-11)} & 2,833 & \(\{A,B,C,G\}\)\\
        SC-63 {\it (LSC-12)} & 2,016 & \(\{A,C\}\)\\
        SC-71 {\it (LSC-12)} & 1,223 & \(\{A,B,C\}\)\\
        SC-84 {\it (LSC-16)} & 217 & \(\{A,G\}\)\\
        \hline
    \end{tabular}
    \end{center}
    
\item Models with 6 or fewer group factors tend towards the origin (LC-01) in the latent layer plot \ref{fig:local_cluster_groups} and are only found in: LC-(01-07), LC-10, LC-12, and LC-13.
\item All models with 1 or 2 group factors are found in SC-083 {\it (LSC-16)} 

\item All 3-cliques are found in: 
    \begin{center}
    \begin{tabular}{c|c|c}
        \hline
        (Sub-)Cluster & \# 3-group models & Group Factor Types \\ \hline
        C-175 {\it (LC-03)} & 1,521 & \(A_{3-11},A_{14-22},B_{3,5-11},C_{2-5},D_{4-11},E_7,G_2\) \\ 
        SC-05 {\it(LSC-01)} & 1 & \(C_2\) \\
        SC-79 {\it (LSC-15)} & 164,130 & \(A_{3-24},B_{3-14},C_{2-7},D_{4-14},E_6,E_7,G_2\) \\ 
        SC-80 {\it (LSC-15)} & 29,332 & \(A_{3-24},B_{3-12},C_{2-7},D_{4-13},E_6,E_7,F_4,G_2\)\\ 
        SC-81 {\it (LSC-15)} & 14,836 & \(A_{3-24},B_{3-12},C_{2-7},D_{4-6},D_{9-12},E_6,F_4,G_2\) \\ 
        SC-83 {\it (LSC-16)} & 937,444& All except \(E_8\)\\ \hline
    \end{tabular}
    \end{center}
    
\item All 4-cliques are found in LC-03, C-044 {\it (LC-04)}, SC-001 and SC-005 {\it (LSC-01)}, LSC-(09-16), LSC-18 and LSC-22.
\item All 14-cliques are found in LC-26, C-104 {\it (LC-31)}, C-039 and C-064 {\it (LC-32)}.
\item All 15-cliques are found in C-104 {\it (LC-31)}
\end{itemize}

Along with Table \ref{tab:CGClusters} and Table \ref{tab:LC1Clusters} this gives us a general sense for where models appear in the latent layer. For instance, the lowest order cliques have a tendency to congregate closer to the origin, near LC-01, while larger cliques tend to form more distinct clusters further away. It also seems that there is no discernable preference for the types of group factors which appear in clusters.

One can look at how similar the Gram matrices are within clusters. For clusters containing models with 3, 4, 14, and 15-group models, we find that the off-diagonal elements of the Gram matrices are very highly correlated within each cluster and differ significantly from other clusters. On the other hand, the diagonal elements tend to vary much more significantly. This would suggest that in these cases, the algorithm has identified the off-diagonal elements of the Gram matrix to be the prominent features. Note that this feature is more pronounced for the clusters containing 14 and 15-group models, rather than 3 and 4-group models.

We have only performed here an initial look into the similarity analysis of the clusters. We expect with more extensive searches, more features will become more prominent, although we leave this for future work. We will see an example of clustering of more complex features when we train the classifier to identify properties of anomaly inflow in section \ref{sec:first-classifier}.

\begin{table}[h]
    \centering
    \begin{tabular}{c|rcrc}
        Local & \multirow{2}{4em}{$\#$ of Elements} & k-Clique Types & Group Sub-type \hspace{0.7cm} & Lie-types \\
        Cluster &  &  & \# : {\footnotesize $\{$A, B, C, D, E, F, G$\}$} & {\footnotesize (A,B,C,D,$\rm E_{6,7,8}$,$\rm F_4$,$\rm G_2$)} \\\hline
        LC-01 & 14,241,455 &\multicolumn{3}{l}{\it Given in Table \ref{tab:LC1Clusters}} \\
        LC-02 & 2,437,669 & 5, 6*, 7, 8, 9 & $202,808: \{5, 0, 0, 1, 0, 0, 0\}$  & $\rm A^*BCDE_{6,7}FG$ \\
        LC-03 & 12,132 & 3, 4* & $2,880:\{2,1,0,0,0,0,1\}$ & $\rm ABCDE_7G^*$ \\
        LC-04 & 49,015 & 4, 7*, 8, 9 & $5,671:\{5,1,0,1,0,0,0\}$ & $\rm A^*BCDE_{6,7}G$ \\
        LC-05 & 1,796,387 & 6, 7*, 8, 9, 10, 11 & $193,775: \{6,1,0,0,0,0,0\}$ & $\rm A^*BCDE_{6,7}FG$\\
        LC-06 & 877,575 & 6*, 7, 8, 9, 10, 11 & $90,774: \{5,0,0,1,0,0,0\}$ & $\rm A^*BCDE_{6,7}FG$ \\
        LC-07 & 925,035 & 6, 7*, 8, 9, 10, 11 & $145,660:\{5,1,0,1,0,0,0\}$ & $\rm A^*BCDE_{6,7}FG$ \\
        LC-08 & 239,827 & 8*, 9, 10 & $54,036:\{6,1,0,1,0,0,0\}$ & A*BCDG \\
        LC-09 & 323,922 & 8*, 9, 10, 11 & $61,149:\{7,1,0,0,0,0,0\}$ & A*BCDG \\
        LC-10 & 14,686 & 6*, 7, 8, 9 & $3,408:\{5,0,0,0,0,0,1\}$ & A*BCDG\\
        LC-11 & 85,869 & 7, 8*, 9, 10, 11 & $8,794:\{7,1,0,0,0,0,0\}$ & $\rm A^*BCDE_7G$ \\
        LC-12 & 77,915 & 6, 8, 9*, 10, 11, 12 & $5,551:\{8,1,0,0,0,0,0\}$ & A*BCDG \\
        LC-13 & 537,548 & 5, 6*, 7, 8, 9 & $36,034:\{4,1,0,1,0,0,0\}$ & $\rm A^*BCDE_{6,7}FG$\\
        LC-14 & 435,411 & 7*, 8, 9, 10, 11, 12 & $128,447:\{7,0,0,0,0,0,0\}$ & $\rm A^*BCDE_7G$\\
        LC-15 & 425,297 & 7*, 8, 9, 10 & $83,036:\{6,0,0,1,0,0,0\}$ & $\rm A^*BCDE_7G$ \\
        LC-16 & 130,885 & 7*, 8, 9, 10 & $13,484:\{5,1,0,0,0,0,1\}$ & $\rm A^*BCDE_7G$ \\
        LC-17 & 603,057 & 7, 8*, 9, 10, 11 & $50,531:\{6,1,0,1,0,0,0\}$ & $\rm A^*BCDE_7G$ \\
        LC-18 & 500,691 & 8*, 9, 10, 11 & $93,448:\{6,1,0,1,0,0,0\}$ & A*BCDG\\
        LC-19 & 224,991 & 7*, 8, 9, 10 & $51,009:\{5,1,0,1,0,0,0\}$ & A*BCDG \\
        LC-20 & 34,669 & 8*, 9, 10 & $4,921:\{6,1,0,1,0,0,0\}$ & A*BCDG \\
        LC-21 & 227,133 & 8*, 9, 10, 11 & $23,213:\{5,2,0,1,0,0,0\}$ & A*BCDG\\
        LC-22 & 32,211 & 9, 10*, 11, 12 & $6,992:\{9,1,0,0,0,0,0\}$ & A*BCDG \\
        LC-23 & 53,526 & 10*, 11, 12, 13 & $9,287:\{8,1,0,1,0,0,0\}$ & A*BCDG \\
        LC-24 & 7,223 & 10*, 11, 12 & $882:\{8,1,0,1,0,0,0\}$ & A*BCD \\
        LC-25 & 8,143 & 10 & $1,499:\{7,2,0,1,0,0,0\}$ & A*BCDG \\
        LC-26 & 72,836 & 10*, 11, 12, 13, 14 & $7,630:\{8,1,0,1,0,0,0\}$ & A*BCDG \\
        LC-27 & 290,546 & 9*, 10, 11, 12, 13 & $28,185:\{7,1,0,1,0,0,0\}$ & A*BCDG \\
        LC-28 & 203,622 & 9*, 10, 11, 12, 13 & $21,049:\{7,1,0,1,0,0,0\}$ & A*BCDG \\
        LC-29 & 460,281 & 7*, 8, 9, 10 & $56,256:\{5,1,0,1,0,0,0\}$ & A*BCDG\\
        LC-30 & 182,998 & 8*, 9, 10 & $26,698:\{7,1,0,0,0,0,0\}$ & $\rm A^*BCDE_7G$ \\
        LC-31 & 409,897 & 7, 8*, 9, 10, 11, & $29,527:\{6,1,0,1,0,0,0\}$ & $\rm A^*BCDE_{6,7}FG$ \\
        & & 12, 13, 14, 15 & & \\
        LC-32 & 98,389 & 10*, 11, 12, 13, 14 & $7,995:\{8,1,0,1,0,0,0\}$ & A*BCDG\\
        LC-33 & 59,158 & 9*, 10, 11 & $7,434:\{8,0,0,1,0,0,0\}$ & A*BCDG \\
        LC-34 & 390,562 & 7*, 8, 9, 10 & $41,622:\{5,1,0,1,0,0,0\}$ & $\rm A^*BCDE_{6,7}FG$\\
        LC-35 & 84,221 & 10*, 11, 12 & $11,770:\{7,2,0,1,0,0,0\}$ & A*BCDG \\
        LC-36 & 13,018 & 10, 11*, 12, 13 & $1,437:\{8,1,0,1,0,0,0\}$ & A*BCDG\\
        LC-37 & 46,596 & 9*, 10, 11 & $7,994:\{7,1,0,1,0,0,0\}$ & A*BCDG \\
        LC-38 & 48,752 & 9*, 10, 11 & $8,614:\{8,1,0,0,0,0,0\}$ & A*BCDG\\
        LC-39 & 74,833 & 9*, 10, 11, 12 & $14,316:\{7,1,0,1,0,0,0\}$ & A*BCDG \\
        LC-40 & 86,319 & 9*, 10, 11 & $15,898:\{6,2,0,1,0,0,0\}$ & A*BCDG\\
        LC-41 & 9,435 & 10*, 11, 12 & $1,856:\{7,2,0,1,0,0,0\}$ & A*BCDG\\
        LC-42 & 34,528 & 10, 11*, 12 & $3,423:\{8,2,0,1,0,0,0\}$ & A*BCD \\
    \end{tabular}
    \caption{An overview of the generic model features appearing in each Cluster. All models with $n_T=0$ appear in cluster C-22.
    }
    \label{tab:CGClusters}
\end{table}

\begin{table}[h]
    \centering
    \begin{tabular}{c|rcrc}
        Local & \multirow{2}{4em}{$\#$ of Elements} & k-Clique Types & Group Sub-type \hspace{0.7cm} & Lie-types \\
        Sub-Cluster &  &  & \# : {\footnotesize $\{$A, B, C, D, E, F, G$\}$} & {\footnotesize (A,B,C,D,$\rm E_{6,7,8}$,$\rm F_4$,$\rm G_2$)} \\\hline
        $\rm \text{LSC-}01^\dagger$ & 5,265,377 & 3, 4, 5*, 6, 7, 8 & $442,628: \{5,0,0,0,0,0,0\}$ & $\rm A^*BCDE_{6,7}FG$ \\
        LSC-02 & 909,449 & 6*, 7, 8, 9 & $193,763:\{6,0,0,0,0,0,0\}$ & $\rm A^*BCDE_{6,7}FG$ \\
        LSC-03 & 528,949 & 5*, 6, 7, 8, 9 & $104,342:\{5,0,0,0,0,0,0\}$ & $\rm A^*BCDE_{6,7}G$ \\
        LSC-04 & 642,329 & 5*, 6, 7, 8, 9 & $102,470:\{5,0,0,0,0,0,0\}$ & $\rm A^*BCDE_{6,7}G$ \\
        LSC-05 & 306,632 & 6*, 7, 8, 9, 10, 11 & $74,030:\{6,0,0,0,0,0,0\}$ & $\rm A^*BCDE_{7}G$ \\
        LSC-06 & 120,056 & 6, 7*, 8 & $8,745:\{4,1,0,1,0,0,1\}$ & $\rm A^*BCDE_{6,7}FG$ \\
        LSC-07 & 818,394 & 6*, 7, 8, 9 & $104,224:\{5,1,0,0,0,0,0\}$ & $\rm A^*BCDE_{6,7}FG$ \\
        LSC-08 & 132,950 & 5*, 6, 7 & $22,467:\{4,1,0,0,0,0,0\}$ & $\rm A^*BCDE_{6,7}FG$ \\
        LSC-09 & 484,998 & 4, 5*, 6, 7 & $65,273:\{4,1,0,0,0,0,0\}$ & $\rm A^*BCDE_{6,7}FG$ \\
        LSC-10 & 63,786 & 4*, 5, 6 & $9,325:\{3,0,0,0,0,0,1\}$ & $\rm A^*BCDE_{7}G$ \\
        LSC-11 & 25,053 & 4 & $4,735:\{2,0,0,1,0,0,1\}$ & $\rm A^*BCDE_{6,7}FG$ \\
        LSC-12 & 63,528 & 4*, 5, 6 & $10,260:\{3,0,0,0,1,0,0\}$ & $\rm A^*BCDE_{6,7}FG$ \\
        LSC-13 & 500,231 & 4*, 5, 6 & $63,009:\{4,0,0,0,0,0,0\}$ & $\rm A^*BCDE_{7}G$ \\
        LSC-14 & 577,514 & 4*, 5, 6 & $86,827:\{3,0,0,1,0,0,0\}$ & $\rm A^*BCDG$ \\
        $\rm \text{LSC-}15^\dagger$ & 881,820 & 3, 4*, 5, 6 & $126,583:\{3,0,0,1,0,0,0\}$ & $\rm A^*BCDE_{6,7}FG$ \\
        $\rm \text{LSC-}16^\dagger$ & 1,676,688 & 1, 2, 3*, 4, 5, 6, 7 & $204,985:\{4,0,0,0,0,0,0\}$ & $\rm A^*BCDE_{6,7,8}FG$ \\
        LSC-17 & 497,247 & 5*, 6, 7, 8 & $38,155:\{4,0,0,1,0,0,0\}$ & $\rm A^*BCDE_{6,7}FG$ \\
        LSC-18 & 193,634 & 4*, 5, 6 & $17,656:\{3,0,0,1,0,0,0\}$ & $\rm A^*BCDE_{6,7}FG$ \\
        LSC-19 & 492,197 & 5*, 6, 7 & $40,178:\{3,1,0,1,0,0,0\}$ & $\rm A^*BCDE_{6,7}FG$ \\
        LSC-20 & 34,177 & 5, 6*, 7 & $2,464:\{4,0,0,1,0,0,1\}$ & $\rm A^*BCDE_{6,7}FG$ \\
        LSC-21 & 15,069 & 5*, 6, 7, 8 & $703:\{3,1,0,0,0,0,1\}$ & $\rm A^*BCDE_{6,7}FG$ \\
        LSC-22 & 11,377 & 4*, 5 & $1,339:\{2,1,0,0,0,0,1\}$ & $\rm A^*BCDE_{6,7}FG$\\
    \end{tabular}
    \caption{Overview of Local Cluster 1 (LC-01) from Table \ref{tab:CGClusters}. Local Sub-Clusters (LSC) are grouped by hand after applying {\it hdbscan} with a much smaller cutoff value giving many smaller clusters.\\ ${}^*$ Most frequent type of feature in a given local cluster.\\ ${}^\dagger$ Local Sub-Clusters with $n_T=0$.}
    \label{tab:LC1Clusters}
\end{table}

\subsection{Detection of `peculiar models'}

Beyond providing a method of classifying the space of 6-dimensional supergravities, autoencoders can be used to identify models which are distinctly different from the average. In the ML community, these techniques are known as anomaly detection as they can be used to identify anomalous data. To avoid confusion with the notion of an anomaly in high energy physics, we will instead refer to these models as `peculiar'.

We can use the autoencoder to identify peculiar models by looking for data which it struggles to reconstruct. If this is the case then it must be distinct, somehow, from the usual inputs it is trained on. The accuracy of the reconstruction of data after it has been through the bottleneck in the NN is measured by the loss function. Since the autoencoder minimises the loss function on the training data, one would expect the loss to be low on generic inputs. If the loss is high then the model must be different from generic inputs, indicating that this model may have peculiar features.

\begin{figure}[h!]
    \centering
    \includegraphics[width=0.75\linewidth]{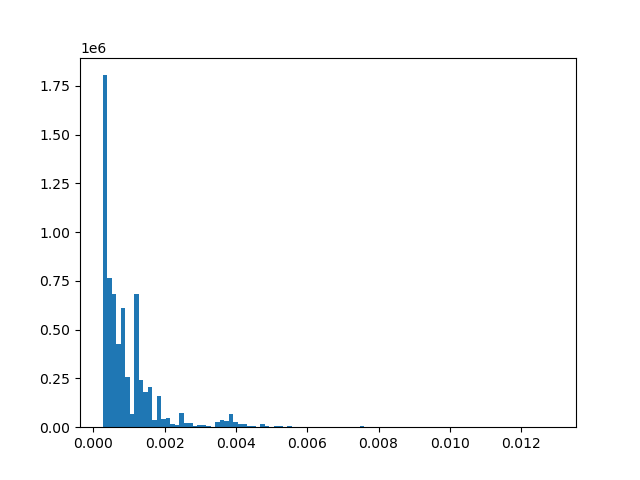}
    \caption{A histogram showing the frequency of different values of the loss function for a sample of inputs.}
    \label{fig:loss_function_hist}
\end{figure}

Plotting a frequency histogram for the loss function for a sample of inputs, we get the graph shown in Figure~\ref{fig:loss_function_hist}.
As expected, the loss is generically very low (although not exactly 0) with an average loss of $2.80358 \times 10^{-4}$. There are, however, a few models with relatively high losses. The largest value of the loss any model has is 0.012908, with the next highest at 0.011349. Given their relatively high value for the loss (over $40\times$ the average loss) we might expect these models to have peculiar properties. Of the models with the highest loss value, most were models with 15 simple gauge group factors. This is unsurprising given that such models make up a particularly small proportion of all models (approximately 0.001\% of the models have 15 simple gauge group factors.). Given that the autoencoder has relatively few examples to train on, it is unsurprising that these models would have the highest loss values.

There is one particularly notable exception to this, however. The fourth highest loss value was 0.010987 and was obtained by two different models (with the same Gram matrix) with only 6 simple gauge group factors. A table showing some of the properties of one of the two models, including the gauge group, representation content, the Gram matrix, and the clique structure, are given below.
\begin{equation*}
\arraycolsep = 5pt
\begin{array}{c|c|c}
\hline\hline
    \text{C-139} &\multicolumn{2}{c}{
    \{\text{A05},\text{A06},\text{A07},\text{A08},\text{A04},\text{G02},\text{A04}\}}\\[0.5ex] \hline
    \begin{matrix}  
        \text{MSLE: }& 0.010987\\
        \text{Coord:}&\{\text{2.14, -4.12}\}
    \end{matrix}
    & \multicolumn{2}{c}{\rm n_H-n_V + 28 \lambda_- = 252,\ \ [\lambda _+,\lambda _-] = [1, 5],\ \  T_{\text{min}} = 6}\\[1ex] \hline\hline
    \text{Vertex Graph}& H^{\rm charged} & \mathbb{G} \\ \hline
    \begin{minipage}{0.3\textwidth}
    \includegraphics[width=\linewidth]{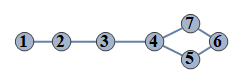}
    \end{minipage}
    &{ \footnotesize
    \begin{matrix}
        \text{(1)  5$\times$(A5-6)},\\ \text{(5)  2$\times$(A4-10)},\\ \text{(7)  2$\times$(A4-10)},\\ \text{(1,2)  1$\times$(A5-6, A6-7)},\\ \text{(2,3)  1$\times$(A6-7, A7-8)},\\ \text{(3,4)  1$\times$(A7-8, A8-9)},\\ \text{(4,5)  1$\times$(A8-9, A4-5)},\\ \text{(5,6)  1$\times$(A4-5, G2-7)},\\ \text{(6,7)  1$\times$(G2-7, A4-5)},\\ \text{(4,7)  1$\times$(A8-9, A4-5)} &   
    \end{matrix}} 
    & {
    \begin{pmatrix}
         9-n_{T} & 0 & 0 & 0 & 0 & 2 & 2 & 2 \\
         0 & -2 & 1 & 0 & 0 & 0 & 0 & 0 \\
         0 & 1 & -2 & 1 & 0 & 0 & 0 & 0 \\
         0 & 0 & 1 & -2 & 1 & 0 & 0 & 0 \\
         0 & 0 & 0 & 1 & -2 & 1 & 0 & 1 \\
         2 & 0 & 0 & 0 & 1 & 0 & 2 & 0 \\
         2 & 0 & 0 & 0 & 0 & 2 & 0 & 2 \\
         2 & 0 & 0 & 0 & 1 & 0 & 2 & 0 \\
    \end{pmatrix}}
\end{array}
\end{equation*}
It is surprising that this model should be so difficult for the autoencoder to reconstruct given its relative simplicity. It is therefore likely that this model has properties that are quite different from the majority of other models. We will not immediately know which differences the algorithm has detected but we can use it as a sorting method to identify models for further study.

Examining this model we note that it is not, on its own, an anomaly-free model since
\begin{equation}
    \Delta = \dim H^{\text{charged}} - n_{V} = 112 > 273 - 29\, T_{\text{min}}
\end{equation}
To form an anomaly-free model, it must be trivially coupled to a model with $\Delta < 0$, $\lambda_{+} = 0$ and a $T_{\text{min}}$ sufficiently small such that the $\tr R^{4}$ anomaly cancels. There are a total of 60 models in the dataset with these properties which correspond to 7 distinct Gram matrices.

Just because a model satisfies $\Delta < 0$, $\lambda_{+} = 0$ and $T_{\text{min}}$ small enough does not guarantee that it will form an anomaly-free model when trivially coupled to the model above. We still require that the anomaly polynomial factorises which requires that the combined Gram matrix $\bbG$ has at most one positive eigenvalue. Coupling the model above with each of the identified 60 models, we find that $\lambda_{+}(\bbG) = 2$ for all admissible $n_{T}$ in every case. Hence, none of these models are fully anomaly free. This does not rule out the possibility that a fully anomaly-free model can be formed by trivially coupling multiple models with $\Delta < 0$, $\lambda_{+} = 0$ and $T_{\text{min}}$ small (indeed we will find such an example below), however it seems to indicate that such combinations are rare.

Most of the 60 possible models have the property
\begin{equation}
    \tilde{\Delta} := \Delta + 29 \lambda_{-} > 0
\end{equation}
When trivially coupled together, the value of $\tilde{\Delta}$ for the total model is just the sum of the $\tilde{\Delta}$ for each of the individual models. Since $T_{\text{min}} \geq \lambda_{-}$, there is only a finite number of possible combinations of such models before the total $\tilde{\Delta} >273$ and hence can't be used to form an anomaly-free model with the high-loss model above. However, there are 8 models with $\tilde{\Delta} < 0$ and hence can be coupled an arbitrary number of times without violating $\tilde{\Delta} \leq 273$ (indeed, they lead to infinite families of anomaly-free theories). We focus on just one of these models, given by the gauge group $E_{6}$ with no charged hypermultiplet matter, and consider trivially coupling $N$ copies to the model above. We find that the smallest value of $N$ and $n_{T}$ for which an anomaly-free model exists is
\begin{equation}
    N = 21 \ , \qquad n_{T} = 62
\end{equation}
further highlighting the difficulty of combining the high-loss model into an anomaly-free model.

We should point out that the autoencoder has only been trained on information contained in the Gram matrix and has not seen further information about the value of $\Delta$, how to trivially combine models, nor the constraint
\begin{equation}
    \Delta + 29 n_{T} \leq 273 \ .
\end{equation}
Nonetheless, these features seem to have been identified and models which do not easily combine with others are identified as `peculiar' by the autoencoder - that is the reconstruction loss for these models is higher than others.

\section{A classifier for probe brane consistency}\label{sec:numerical_labelling}

Our second application of neural networks to the 6d supergravity landscape will be to build a classifier with the aim of determining anomaly-free solutions.  In order to do so, we took a small portion of the data and labeled it as described below.  Using the Gram matrix data as input, we trained a network to reproduce this labeling.

\subsection{A numerical method for labelling the data}

Building a NN which will classify models based on whether they are consistent under the insertion of probe branes requires a large dataset of models with this property labelled. While significant efforts have been made into studying which models satisfy these constraints \cite{Kim:2019vuc,Lee:2019skh,Angelantonj:2020pyr,Cheng:2021zjh,Tarazi:2021duw,Hamada:2024oap}, in each case only models with specific properties were studied. To improve the chances of obtaining a highly accurate classifier, we required training data which was representative of the entire population of supergravity models found in \cite{GLogesData1,GLogesData2}. To ensure that, we took a large random sample of the full dataset (of $\mathcal{O}(10^{5})$ data points) and designed a procedure to label this set.

The question of consistency under probe branes is a problem in non-linear integer programming. That is, we need to check the positivity of the following non-linear function
\begin{equation}\label{eq:objective_function}
    f(Q) = c_{l} - \sum_{i}\frac{k_{i}\dim G_{i}}{k_{i} + \hat{h}_{i}}
\end{equation}
over points in a lattice $Q\in \Lambda_{S}$. The points in the lattice must also satisfy non-linear constraints. Such problems are computationally very expensive. Even finding the appropriate decomposition of the Gram matrix \eqref{eq:local_anom_freedom} for $\bbD$ an integer matrix can be challenging when $n_{T}$ is large and makes the problem intractable when trying to perform this calculation on $\mathcal{O}(10^{5})$ data points. In order to make progress, we will make some simplifications of the problem which we outline in this section. These simplifications will affect the accuracy of our labeling as we shall discuss.

The first simplification that we make is, rather than working over points $Q$ in the lattice, we work with the $\Orth{1,n_{T}}$ invariant combinations, $Q\cdot Q$, $Q\cdot b_{i} = k_{i}$, and $Q\cdot a$. For convenience, we will write $k_{0} = Q\cdot a$ and $q = Q\cdot Q$. Working over these parameters has the benefit of not only being $\Orth{1,n_{T}}$ invariant, but also simplifying the non-linear constraints on $Q$ to linear constraints on $q,k_{0},k_{i}$. Indeed, the constraints given by \eqref{eq:inflow_cond_1} simplify to
\begin{align}
    0\leq c_{l} &= 3q + 9k_{0} + 2 \\
    0 \leq c_{r} &= 3q+3k_{0} \\
    0 \leq k_{l} & =\tfrac{1}{2}(q-k_{0} + 2) \\
    0 \leq k_{i} &
\end{align}
Since the charge lattice is integral, we also need $q,k_{0},k_{i} \in \bbZ$. Note that we will ignore the positive tension condition here as well.

We can also simplify the problem by considering only the values of $Q$ (or $q, k_{0}, k_{i}$) for which $c_{l}$ takes small values. This will generically provide the strongest constraints on the possible groups $G_{i}$ and the representation content, which appears in \eqref{eq:objective_function} through the $b_{i}$ and hence $k_{i}$. It is clear from the expressions above that $c_{l}$ takes small values when both $q$ and $k_{0}$ are small. We therefore consider only the following solutions to the above inequalities
\begin{equation}\label{eq:allowed_charges}
    (q,k_{0}) = (-1,1),\, (0,0),\, (0,2), \, (1,1), \, (1,3), \, (2,0),\, (2,2), \, (2,4)
\end{equation}
Since $a$ is a characteristic element of the charge lattice (See \eqref{eq:characteristic}), $q$ and $k_{0}$ are either both odd or both even. At the above values, the left-moving central charge takes the following values, respectively:
\begin{equation}
    c_{l} = 8, 2, 20, 14, 32, 8, 26, 44.
\end{equation}
By considering these different values of $(q,k_{0})$, we end up with 8 different constraints on the $G_{i}$ and $k_{i}$ from requiring that \eqref{eq:objective_function} is positive. Note that satisfying the positivity of \eqref{eq:objective_function} for each of these values of $c_{l}$ is necessary for consistency under anomaly inflow, but not sufficient. We shall refer to this necessary condition as the \emph{restricted} anomaly inflow condition.

One may wonder why we can't simply take the value of $(q,k_{0})$ with the smallest $c_{l}$, i.e. $(0,0)$, since that would seem to give the strongest constraint on the groups and levels. However, since $q,k_{0},k_{i}$ are all defined in terms of $Q$, they are not all independent. For example, if $n_{T} < 9$, then the only value of $Q$ which gives $(q,k_{0}) = (0,0)$ is $Q=0$. In this case, we get $k_{i} = 0$ and so we actually get no constraints on the groups $G_{i}$. More generally, we need to understand how the values $q,k_{0}, k_{i}$ are related to understand how the bounds constrain the groups.

Before we do that, we make another simplification. We relax the condition $k_{i} \in \bbZ$ and instead allow for $k_{i}\in \bbR$. At the level of the charge lattice, this is equivalent to considering $Q$ as an element of the vector space spanned by the charge lattice $\Lambda_{S}\otimes \bbR$. The benefit of this approach is that we do not need to have detailed knowledge of the charge lattice, nor of the integer matrix $\bbD$ in the decomposition of $\bbG$ which, as we discussed earlier, is computationally expensive to determine. Instead, we are free to utilise the $\Orth{1,n_{T}}$ invariance of the problem to find a convenient representation of $\bbD$. One particularly convenient choice for $\bbD$ is the following. Since $\bbG$ is a symmetric matrix, there is some orthonormal basis of eigenvectors which we write collectively as $X$, with a diagonal matrix of eigenvalues $\Lambda$. Then
\begin{equation}
    \bbG = X\Lambda X^{T} \qquad \Rightarrow \qquad \bbD = X |\Lambda|^{\tfrac{1}{2}}
\end{equation}
where $|\Lambda|$ is a diagonal matrix with the absolute values of the eigenvalues down the diagonal. Note that we have implicitly ordered the eigenvectors and eigenvalues so that the positive eigenvalue of $\bbG$ (if it exists) comes first. If $\lambda_{+}(\bbG)= 0$, we pad $\bbD$ on the left with a single 0. If $\lambda_{-}(\bbG) < n_{T}$ we pad $\bbD$ on the right with 0's so that it has dimension $n\times (n_{T}+1)$. This representation of $\bbD$ will be some $\Orth{1,n_{T}}$ transformation of the integer matrix whose rows lie in the charge lattice $\Lambda_{S}$ (which, through an abuse of notation, we also write as $\bbD$). This also has the benefit of satisfying the following
\begin{equation}
    \bbD^{T}\bbD = |\Lambda|^{\tfrac{1}{2}} X^{T} X|\Lambda|^{\tfrac{1}{2}} = |\Lambda|^{\tfrac{1}{2}} I|\Lambda|^{\tfrac{1}{2}} = |\Lambda|
\end{equation}
which simplifies many calculations. Since there is some $\Orth{1,n_{T}}$ frame in which $\bbD^{T}\bbD$ is diagonal, this also implies that it, and its pseudo-inverse $(\bbD^{T}\bbD)^{-1}$ (see below) commute with $\eta$.

The downside of this simplification is that, any conclusion we make using a particular value of the $k_{i}$ cannot be directly related to properties of models with a probe brane inserted. This is because we cannot be certain that there is some dyonic string with charge $Q\in \Lambda _{S}\subset \Lambda_{S}\otimes \bbR$ for which $\bbD\eta Q= k$. To resolve this issue, we can demand that the function \eqref{eq:objective_function} is positive for \emph{all} allowed values of $k_{i}$, up to the constraints above, at all specified values of $(q,k_{0})$. Certainly, this is a sufficient condition to pass the inflow consistency conditions, but is not necessary. We therefore expect our procedure to miss many consistent models. However, when it identifies a model we can be certain that it is consistent under insertion of probe branes with charge such that $(q,k_{0})$ have the specified values.

Here, and in the rest of this section, the superscript $-1$ will denote the `pseudo-inverse' of symmetric matrices. That is, for some symmetric matrix $X$, with eigenvalues $\lambda_{i}(X)$ \footnote{These should not be confused with either $\lambda_\pm(\mathbb{G})$ which are the number of positive/negative eigenvalues of $\mathbb{G}.$ That is, $\lambda_+(\mathbb{G}) = {\rm dim}(\{\lambda_i| \lambda_i >0\})\leq1$ and $\lambda_-(\mathbb{G}) = {\rm dim}(\{\lambda_i| \lambda_i <0\})\leq {\rm n_T}$.} and eigenvectors $e_{i}$, the pseudo-inverse $X^{-1}$ is defined to have the same eigenvectors $e_{i}$, with eigenvalues
\begin{equation}
    \lambda_{i}(X^{-1}) = \begin{cases}
        \lambda_{i}(X)^{-1} & \lambda_{i}\neq 0 \\
        0 & \lambda_{i} = 0
    \end{cases}
\end{equation}
Note that this implies that $X^{-1}$ has the same kernel as $X$ and $XX^{-1} = X^{-1}X$ is the projector onto the image of $X$.

With these observations and simplifications, we return to finding the relationship between $q,\, k_{0},\, k_{i}$. To do so, we write the levels in a vector $k = (k_{0},\, k_{1},\, ...,\, k_{n})$. Then, the relation between $Q$ and $k$ is
\begin{equation}
    \bbD\,\eta\, Q = k \ , \qquad \bbG = \bbD\,\eta\, \bbD^{T} \ 
\end{equation}
where $\bbG$ is the Gram matrix of the model and $\bbD$ is the decomposition matrix required for local anomaly cancellation, as in \eqref{eq:local_anom_freedom}. From this, it is clear that the $k_{0},k_{i}$ may not be independent, but instead must be in the image of $\bbD\eta$. We can define an orthogonal projector onto the image of $\bbD\eta$ (and hence onto the image of both $\bbD$ and $\bbG$) by
\begin{equation}
    P = \bbD(\bbD^{T}\bbD)^{-1}\bbD^{T}
\end{equation}
The first constraint on the level vector $k$ is then
\begin{equation}
    Pk = k
\end{equation}
or equivalently
\begin{equation}
    ||Pk - k||_{2} = (Pk - k)^{T}(Pk - k) = k^{T}k - k^{T}Pk = 0.
\end{equation}
Assuming this to be the case, we can find a Q which maps to $k$ via the following.
\begin{equation}
    Q = \eta(\bbD^{T}\bbD)^{-1} \bbD^{T} k
\end{equation}
If $\bbD$ is not full rank, i.e. if $\lambda_{+}(\bbG) = 0$ or $\lambda_{-}(\bbG) < n_{T}$, then the map $k=\bbD\eta Q$ has a kernel and the most general value of $Q$ is
\begin{equation}
    Q = \eta(\bbD^{T}\bbD)^{-1} \bbD^{T} k + x \ , \quad \bbD\,\eta\, x = 0 \ .
\end{equation}
From this, we see that the value of $q = Q\cdot Q$ is given by
\begin{align}
    \begin{split}\label{eq:q_equation}
        q &= Q^{T} \,\eta \, Q \\
        &= (\eta(\bbD^{T}\bbD)^{-1} \bbD^{T} k + x)^{T}\, \eta \,  (\eta(\bbD^{T}\bbD)^{-1} \bbD^{T} k + x) \\
        &= k^{T} \bbD (\bbD^{T}\bbD)^{-1} \eta (\bbD^{T}\bbD)^{-1} \bbD^{T} k + x^{T} (\bbD^{T}\bbD)^{-1} \bbD^{T} k + k^{T} \bbD (\bbD^{T}\bbD)^{-1} x + x^{T} \eta x \\
        &= k^{T} \bbD (\bbD^{T}\bbD)^{-1} \eta (\bbD^{T}\bbD)^{-1} \bbD^{T} k + 2k^{T} \bbD (\bbD^{T}\bbD)^{-1} x + x^{T} \eta x.
    \end{split}
\end{align}
Let's consider the final line term by term. 

For the first term, note that
\begin{equation}
    \begin{split}
        \bbD (\bbD^{T}\bbD)^{-1} \eta (\bbD^{T}\bbD)^{-1} \bbD^{T} \bbG &= \bbD (\bbD^{T}\bbD)^{-1} \eta (\bbD^{T}\bbD)^{-1} \bbD^{T} \bbD \eta \bbD^{T} \\
        &= \bbD(\bbD^{T}\bbD)^{-1} \bbD^{T} \\
        &= P \\
        \Rightarrow \quad \bbD (\bbD^{T}\bbD)^{-1} \eta (\bbD^{T}\bbD)^{-1} \bbD^{T} &= \bbG^{-1}
    \end{split}
\end{equation}
where in the final line we have used the fact that $P$ is the projector onto the image of $\bbG$, and hence this satisfies the condition for pseudo-inverse. Hence, the first term in \eqref{eq:q_equation} is simply given by $k^{T}\bbG^{-1} k$.

For the second term, notice that
\begin{equation}
\begin{aligned}
    2k^{T} \bbD (\bbD^{T}\bbD)^{-1} x &= 2k^{T} \bbD (\bbD^{T}\bbD)^{-1} (\bbD^{T}\bbD)^{-1} (\bbD^{T}\bbD) \eta^{2} x \\
    &= 2k^{T} \bbD (\bbD^{T}\bbD)^{-1} (\bbD^{T}\bbD)^{-1} \eta \bbD^{T}(\bbD \eta x) \\
    &= 0.
\end{aligned}
\end{equation}
In the second line we have used the fact that $\bbD^{T}\bbD$ commute with $\eta$. Hence, we have
\begin{equation}
    q = k^{T}\bbG^{-1}k + x\cdot x.
\end{equation}

As we will discuss in more detail later, if $\lambda_{+}(\bbG) = 0$ and $\lambda_{-}(\bbG)<n_{T}$ then there is an ambiguity in the definition of $\bbD$. This leads to an ambiguity in determining whether the model is consistent under anomaly inflow and so we will not consider such cases. Assuming that we do not have both $\lambda_{+}(\bbG) = 0$ and $\lambda_{-}(\bbG) < n_{T}$, then we have 3 possible cases.
\begin{equation}
\arraycolsep = 2pt
\begin{array}{rlrl}
    \lambda_{+}(\bbG) &= 0 &\quad x\cdot x \geq 0 & \quad q\geq k^{T}\bbG^{-1}k \\[5pt]
    \lambda_{-}(\bbG) &< n_{T} &\quad x\cdot x \leq 0 & \quad q\leq k^{T}\bbG^{-1} k \\[5pt]
    \lambda_{+}(\bbG) = 1 , \ \lambda_{-}(\bbG) &= n_{T} & \quad x = 0 & \quad  q = k^{T}\bbG^{-1} k
\end{array}
\end{equation}
We see that, depending on the eigenvalues of $\bbG$ and the value of $n_{T}$, the value of $q$ is bounded below or above, or is equal to $k^{T}\bbG^{-1}k$. 

To summarise, our labelling procedure runs as follows. Using minimisation commands in the `scipy' package in python, we maximise the following function over $k_{i}\in \bbR$ subject to the following constraints.
\begin{equation}\label{eq:label_procedure_1}
    \begin{array}{ll}
        \text{maximise:} \qquad & \tilde{f}(k) = \sum_{i>0} \frac{k_{i}\dim G_{i}}{k_{i} +\check{h}_{i}} \\[20pt]
        \text{subject to:} & \begin{array}{rl}
            0 &= k^{T}k - k^{T}Pk
            \end{array}\\[5pt]
        & \begin{array}{rl}   
            q &\left\{  \begin{array}{ll}
                \geq k^{T}\bbG^{-1}k & \quad \lambda_{+}(\bbG) = 0  \\
                \leq k^{T}\bbG^{-1}k & \quad \lambda_{-}(\bbG) < n_{T} \\
                = k^{T}\bbG^{-1}k & \quad \lambda_{+}(\bbG) = 1,\ \lambda_{-}(\bbG) = n_{T}
            \end{array} \right.
        \end{array}\\[20pt]
        & (q,k_{0}) \text{ fixed} \\[5pt]
        & 0\leq k_{i}
    \end{array}
\end{equation}
If the maximum value of $\tilde{f}$ is less than the corresponding value of $c_{l}$ for each specified pair of values $(q,k_{0})$, then we label the model 0. Otherwise, we label the model with a 1. Note that, in the labelling process, we took the value of $n_{T}$ to be the minimum value $T_{\text{min}}$ specified in the dataset. We also labelled any model with $\lambda_{+}(\bbG) = 0$ and $\lambda_{-}(\bbG)<n_{T}$ with a 1. This is because the ambiguity in the definition of $\bbD$, mentioned earlier and discussed in more detail later, makes determining the consistency of these models tricky. We would like our 0 labels to be as accurate as possible and hence we disregarded these models, labelling them 1. Finally, there may be some models and some values of $(q,k_{0})$ for which there are no values of $k_{i}$ satisfying the above constraints. When this occurs, the positivity of \eqref{eq:objective_function} is vacuously satisfied.

When labelling the data using the above method, we found an incidence rate of 0's of around 2.5\%. While this still allowed us to find many thousands of 0's when labelling $\mathcal{O}(10^{5})$ models, the huge bias in the dataset between 0's and 1's would make building an effective classifier challenging. We will discuss this more in the following sections.

To find a more balanced dataset, we also decided to use a different labelling process on the same sample. Instead of finding a process which would be effective at finding models which were consistent under the restricted inflow criteria, we designed a method which would be more effective at identifying models which were inconsistent under the insertion of probe branes. That is, we looked for models for which \eqref{eq:objective_function} is negative for some values of $Q$. To do so, we made the same simplifications as above but instead minimised the function $\tilde{f}(k)$ subject to the same constraints, as in \eqref{eq:label_procedure_1}. If, for some value of $(q,k_{0})$, the minimum is \emph{greater} than the corresponding value of $c_{l}$, then we know that the model will be inconsistent for \emph{any} probe brane with charge $Q\in \Lambda_{S}$ which give $(q,k_{0})$ and satisfy the constraints. The problem with this method is that it is not a necessary condition, since models may have \eqref{eq:objective_function} positive or negative for different values of $Q$ giving the same values of $(q,k_{0})$. It is also not a sufficient condition. To see this, we note that we may have
\begin{align}
    \{Q\in \Lambda_{S}\otimes\bbR\,|\, q,k_{0} \text{ fixed},\, k_{i}\geq 0\} &\neq \emptyset\\
    \{Q\in \Lambda_{S}\,|\, q,k_{0} \text{ fixed},\, k_{i}\geq 0\} &= \emptyset
\end{align}
Hence, while the value of \eqref{eq:objective_function} is negative for all $Q$ in the first set, there is no dyonic string that one can insert which would give the same values of $q,k_{0},k_{i}$, which would be given by the second set above.

Nonetheless, the property that \eqref{eq:objective_function} is negative for all $Q\in \Lambda_{S}\otimes \bbR$ subject to the constraints is a strong indication that it is in the swampland. We therefore used the following labelling procedure to find a second labelled set.
\begin{equation}\label{eq:label_procedure_2}
    \begin{array}{ll}
        \text{minimise:} \qquad & \tilde{f}(k) = \sum_{i>0} \frac{k_{i}\dim G_{i}}{k_{i} +\check{h}_{i}} \\[20pt]
        \text{subject to:} & \begin{array}{rl}
            0 &= k^{T}k - k^{T}Pk
            \end{array}\\[5pt]
        & \begin{array}{rl}   
            q &\left\{  \begin{array}{ll}
                \geq k^{T}\bbG^{-1}k & \quad \lambda_{+}(\bbG) = 0  \\
                \leq k^{T}\bbG^{-1}k & \quad \lambda_{-}(\bbG) < n_{T} \\
                = k^{T}\bbG^{-1}k & \quad \lambda_{+}(\bbG) = 1,\ \lambda_{-}(\bbG) = n_{T}
            \end{array} \right.
        \end{array}\\[20pt]
        & (q,k_{0}) \text{ fixed} \\[5pt]
        & 0\leq k_{i}
    \end{array}
\end{equation}
If the minimum value of $\tilde{f}$ is greater than $c_{l}$ then we label the model with a 1. Otherwise, we label it with a 0. Again, in the labelling process, we assumed that the value of $n_{T}$ was the minimum value $T_{\text{min}}$ specified in the dataset. We also labelled the ambiguous models with $\lambda_{+}(\bbG) = 0$ and $\lambda_{-}(\bbG)<n_{T}$ with a 0.

When labelling the data through this second process, we found an incidence rate of 1's of about 22\%. While this still implies an imbalanced dataset, it is significantly more balanced than the first. We summarise the two datasets below.
\begin{enumerate}
    \item\label{data1} The first dataset was labelled through the procedure in \eqref{eq:label_procedure_1}. Its total size is 546,116 models, of which 13,649 are 0's and 532,467 are 1's. If a model is labelled with a 0 in this dataset, then it is consistent under the insertion of probe branes with charge $Q$ with corresponding values of $(q,k_{0})$ as in \eqref{eq:allowed_charges}.

    \item\label{data2} The second dataset was labelled through the procedure in \eqref{eq:label_procedure_2}. Its total size is 111,457 models, of which 88,401 are 0's and 23,056 are 1's.
\end{enumerate}

Note that we have chosen to label only the building blocks of anomaly free models that were enumerated in \cite{Hamada:2023zol}. These models can be combined together trivially\footnote{Trivial in the sense that there is no cross-representation content between the different building blocks.} to form new models. Indeed, for some building blocks, it is necessary to do this in order to make the gravitational $\tr R^{4}$ anomaly vanish. Hence we briefly describe how our labelling behaves under trivial combination. When we add models together, we increase the dimension of the Gram matrix $\bbG$ and introduce an additional set of levels and therefore we increase the number of constraints in \eqref{eq:label_procedure_1} and \eqref{eq:label_procedure_2}. As such, when combining models together, one of two things can happen. Either (1) there are no longer any solutions to the constraint equation or (2) there is a solution to the constraint equation but now the function $\tilde{f}$ we want to optimise gets additional positive terms.

Suppose that some model, with gram matrix $\bbG_{1}$ and optimising function $\tilde{f}_{1}$ has been labelled by the second classifier with a 1. That is, everywhere in which the constraints are satisfied we have $\tilde{f}_{1}(k) > c_{l}$. Now suppose we combine it with some other model such that the new theory has Gram matrix and optimising function
\begin{equation}
    \bbG_{3} = \bbG_{1} \oplus \bbG_{2} \ , \qquad \tilde{f}_{3} = \tilde{f}_{1} + \tilde{f}_{2} \ .
\end{equation}
Note here $\oplus$ denotes trivial product in the sense that the sub-matrix $C_{ij}$ in
\eqref{eq:Gram} is block-diagonal. Then, everywhere that the constraints are satisfied, we have
\begin{equation}
    \tilde{f}_{3} \geq \tilde{f}_{1} > c_{l}
\end{equation}
Hence, this model remains inconsistent after combining with other models in the set, provided there are $k_{i}$ which solve the constraints.

On the other hand, for models labelled with a 0 by the first classifier, the condition $\tilde{f} < c_{l}$ everywhere around the origin may be broken as we add more and more positive terms to $\tilde{f}$. So the labelling is not necessarily preserved by combining models together for the first classifier. However, we know that two models labelled with a 0 under the first classifier are guaranteed to combine to a consistent model (in the sense of the restricted anomaly inflow criteria we have described in this section) if
\begin{equation}
    \max \tilde{f}_{1} + \max \tilde{f}_{2} \leq c_{l}
\end{equation}
Hence, the labelling of the first classifier is preserved under certain combinations of models which have been labelled with a 0.

\subsection{Training the Classifier}

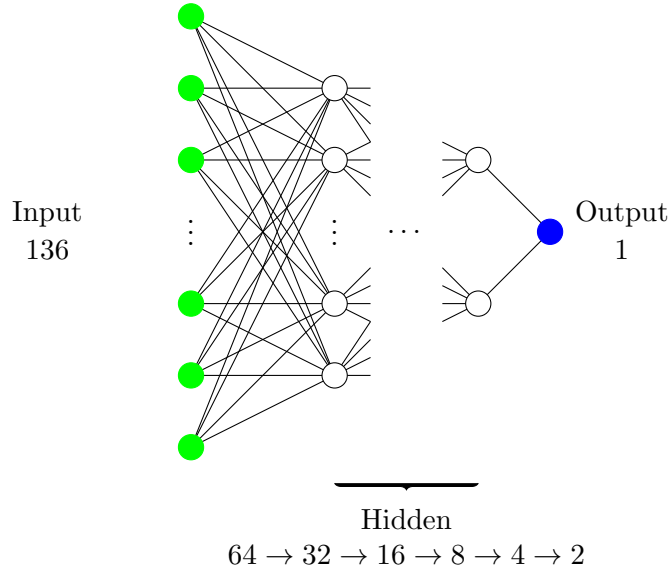
\begin{figure}[h]
    \centering
\begin{tikzpicture}[roundnode/.style={circle, thick, minimum size=7mm}, scale=0.95] 
        \draw (-5,3) -- (-3,2);
        \draw (-5,3) -- (-3,1);
        \draw (-5,3) -- (-3,-1);
        \draw (-5,3) -- (-3,-2);
        \draw (-5,2) -- (-3,2);
        \draw (-5,2) -- (-3,1);
        \draw (-5,2) -- (-3,-1);
        \draw (-5,2) -- (-3,-2);
        \draw (-5,1) -- (-3,2);
        \draw (-5,1) -- (-3,1);
        \draw (-5,1) -- (-3,-1);
        \draw (-5,1) -- (-3,-2);
        \draw (-5,-1) -- (-3,2);
        \draw (-5,-1) -- (-3,1);
        \draw (-5,-1) -- (-3,-1);
        \draw (-5,-1) -- (-3,-2);
        \draw (-5,-2) -- (-3,2);
        \draw (-5,-2) -- (-3,1);
        \draw (-5,-2) -- (-3,-1);
        \draw (-5,-2) -- (-3,-2);
        \draw (-5,-3) -- (-3,2);
        \draw (-5,-3) -- (-3,1);
        \draw (-5,-3) -- (-3,-1);
        \draw (-5,-3) -- (-3,-2);
        \filldraw[color=green, fill=green] (-5,3) circle (5pt);
        \filldraw[color=green, fill=green] (-5,2) circle (5pt);
        \filldraw[color=green, fill=green] (-5,1) circle (5pt);
        \node[roundnode] at (-5,0.1) {$\vdots$};
        \filldraw[color=green, fill=green] (-5,-1) circle (5pt);
        \filldraw[color=green, fill=green] (-5,-2) circle (5pt);
        \filldraw[color=green, fill=green] (-5,-3) circle (5pt);
        \node[roundnode] at (-7,0.25) {Input};
        \node[roundnode] at (-7,-0.25) {136};
        %
        \draw (-3,2) -- (-2.5,2);
        \draw (-3,2) -- (-2.5,1.75);
        \draw (-3,2) -- (-2.5,1.5);
        \draw (-3,2) -- (-2.5,1.25);
        \draw (-3,1) -- (-2.5,1.25);
        \draw (-3,1) -- (-2.5,1);
        \draw (-3,1) -- (-2.5,0.75);
        \draw (-3,1) -- (-2.5,0.5);
        \draw (-3,-1) -- (-2.5,-0.5);
        \draw (-3,-1) -- (-2.5,-0.75);
        \draw (-3,-1) -- (-2.5,-1);
        \draw (-3,-1) -- (-2.5,-1.25);
        \draw (-3,-2) -- (-2.5,-1.25);
        \draw (-3,-2) -- (-2.5,-1.5);
        \draw (-3,-2) -- (-2.5,-1.75);
        \draw (-3,-2) -- (-2.5,-2);
        \node[roundnode] at (-2,0) {$\dots$};
        \filldraw[color=black, fill=white] (-3,2) circle (5pt);
        \filldraw[color=black, fill=white] (-3,1) circle (5pt);
        \node[roundnode] at (-3,0.1) {$\vdots$};
        \filldraw[color=black, fill=white] (-3,-1) circle (5pt);
        \filldraw[color=black, fill=white] (-3,-2) circle (5pt);
        \draw (-1.5,1.25) -- (-1,1);
        \draw (-1.5,1) -- (-1,1);
        \draw (-1.5,0.75) -- (-1,1);
        \draw (-1.5,0.5) -- (-1,1);
        \draw (-1.5,-1.25) -- (-1,-1);
        \draw (-1.5,-1) -- (-1,-1);
        \draw (-1.5,-0.75) -- (-1,-1);
        \draw (-1.5,-0.5) -- (-1,-1);
        \draw (-1,1) -- (0,0);
        \draw (-1,-1) -- (0,0);
        \filldraw[color=black, fill=white] (-1,1) circle (5pt);
        \filldraw[color=black, fill=white] (-1,-1) circle (5pt);
        \filldraw[decoration = {brace, mirror}, decorate] (-3,-3.5) -- (-1,-3.5);
        \node[roundnode] at (-2,-4) {Hidden};
        \node[roundnode] at (-2,-4.5) {$64\to32\to16\to8\to4\to2$};
        \filldraw[color=blue, fill=blue] (0,0) circle (5pt);
        \node[roundnode] at (1,0.25) {Output};
        \node[roundnode] at (1,-0.25) {1};
    \end{tikzpicture}
    
    \caption{A diagram of the classifier used to learn which solutions are anomalous.  The input (\textcolor{green}{green}), as with our earlier autoencoder, was a 16x16 symmetric Gram matrix.  The layers halve in size at each step, resulting in an output (\textcolor{blue}{blue}) which is a single node.  Its sigmoid activation ensures that its output will lie in the range (0,1).}
    \label{fig:classifier}
\end{figure}

Our goal was to train a neural network to reproduce the above labelling of the data.  In contrast to our autoencoder, which was unsupervised, this network is a classic case of supervised learning -- a classifier which takes solutions as input and learns to give their label as an output.  Figure~\ref{fig:classifier} shows the network configuration, which begins with the same 136-dimensional input as the autoencoder.  This allows us to feed in the Gram matrix information.  We considered supplementing this input with the much larger clique vector information carried by each solution, but in testing this resulted in much larger network sizes with no noticeable increase in performance, so we opted for a purely Gram matrix-based implementation.  Successive dense layers in the network contain half the nodes of the previous layer, creating a funnel in which the matrix data is gradually mapped to a single number, in the final dense layer with a single node.  The hidden layers use ReLu outputs to introduce nonlinearity, while the output layer uses a sigmoid activation function.  This guarantees that the network output lies between 0 and 1 -- rounding that output gives us a definite labelling for each Gram matrix we input.

With an architecture defined, we move on to training networks on the two sets of labelled data.  Unlike the autoencoder, which had access to our full set of data, this is a situation where we were only practically able to label a small portion of the data, and must rely on that for training.  Thus, it is important to hold a portion of the data in reserve.  This is known as the validation set, and the network does not see this data during training, but its performance on it is evaluated each epoch.  Performance on the validation set gives us an idea of how this network would perform on any random set of solutions which it hasn't seen, which is important since we want to later apply the trained model to the full dataset.  We also want to present the network with a set of data in which the 0 and 1 labels are balanced.  If, for instance, we showed it the full first dataset in which only 2.5\% of the labels were 0, the network would immediately be drawn towards the simple solution of `label every solution with a 1' as that would provide 97.5\% accuracy, hence low loss.  To avoid this, we undersample the 1 labels to provide a balanced set of 27,298 solutions, half with each label.  Of these, we reserve 4,000 for validation and use the remainder for training.  Now, in order to improve over random guessing, the network must actually take into account features of the input data.  At best, we ended up with a network which achieved 97.96\% accuracy on the training set and 96.72\% accuracy on validation.

We repeated the process on the second set of data, for which the initial imbalance was not as bad: 20.7\% of the 111,457 solutions were labelled with 1.  This gave us a balanced dataset of 46,112 solutions (with the 0 labels chosen randomly at time of training), of which we reserved 6,000 for validation and trained on the remainder.  Here, our best model achieved 88.60\% accuracy on the training set and 85.53\% accuracy on validation.  While it may seem at first glance that the first model is more accurate, the fact that its training set had to be drawn from a more imbalanced set of data will impact how well it performs on a truly random sample (note that the validation set, like the training set, was balanced in its labels, which is unrealistic).  Our main task for these classifiers will be to identify solutions which are non-anomalous, so the relevant statistic is how often it does that incorrectly i.e.~what is its rate of false positives.  If we assume that the true incidence rates of 0 and 1 labels, using the respective labelling methods, match the samples we used here, then the first classifier would give a false positive 57\% of the time, while the second would perform better at a 39.3\%.  Showing the entirety of our data to both models, then, we expect that the second one will correctly predict non-anomalous models more often than the first.  In the following section we will discuss how the naive rounding of the network output can be tweaked to further improve this situation.

\subsection{Putting the Classifier to Work}

We now have two classifiers built from the two different datasets labelled in section \ref{sec:numerical_labelling}. Applying these to the entire set of 6-dimensional supergravity models, we can obtain predictions about which models will satisfy the restricted anomaly inflow criterion, and which will violate it and hence are likely in the swampland. We will examine the two classifiers in turn.

\subsubsection{The first classifier}\label{sec:first-classifier}

The first classifier was trained on dataset \ref{data1}. Recall that this dataset labelled with a 0 any model whose central charge satisfied
\begin{equation}\label{eq:inflow_condition}
    c_{l} - \sum_{i}\frac{k_{i}\dim G_{i}}{k_{i} +\check{h}_{i}} \geq 0
\end{equation}
for all $Q \in \Lambda_{S}\otimes \bbR$ such that
\begin{align}
    (q,k_{0}) &= (-1,1),\, (0,0),\, (0,2), \, (1,1), \, (1,3), \, (2,0),\, (2,2), \, (2,4) \\
    k_{i} &\geq 0 \\
    c_{l},c_{r}, k_{l} &\geq 0
\end{align}
In this dataset, the points of interest are the 0's since these represent models that pass the restricted anomaly inflow criteria for all points $Q$ in the charge lattice $\Lambda_{S}$ which satisfy the above constraints. In contrast, if a model is labelled with a 1 it does not guarantee that it fails the anomaly inflow test \eqref{eq:inflow_condition} at some point $Q\in \Lambda_{S}$, but instead at some point $Q\in \Lambda_{S}\otimes \bbR$ (a necessary but not sufficient condition). 

\begin{figure}
    \centering
    \begin{subfigure}{0.45\textwidth}
        \includegraphics[width=0.95\linewidth]{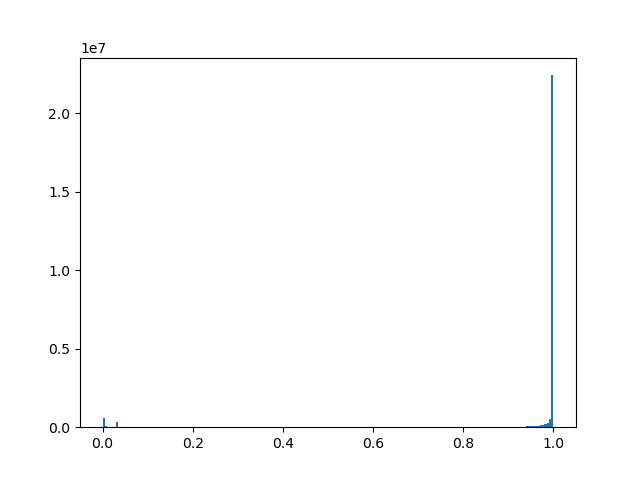}
    \end{subfigure}
    \begin{subfigure}{0.45\textwidth}
        \includegraphics[width=0.95\linewidth]{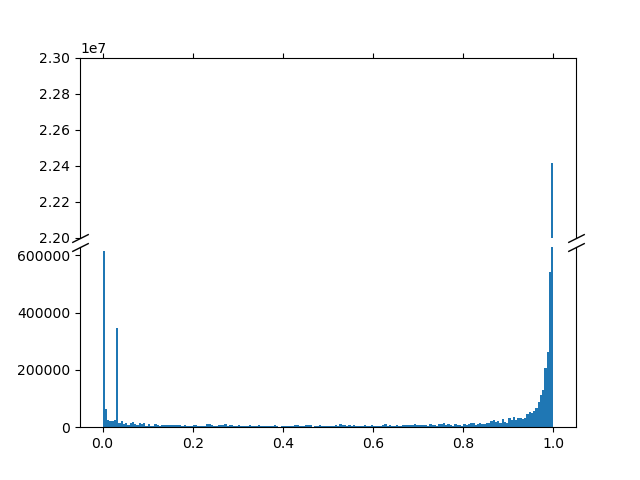}
    \end{subfigure}
    \caption{A frequency histogram of the outputs of the classifier. We see that there are significantly more values close to 1 than close to 0. This matches the training dataset which had around 2.5\% of 0's. The histogram on the right has the y-axis cut to better see the frequency of models with a prediction of close to 0.}
    \label{fig:classifier_1}
\end{figure}

This classifier had a validation accuracy of 96.7\%. Applying this classifier to the full dataset, we get the histogram of outputs as in Figure \ref{fig:classifier_1}. Note that there are two large peaks, one close to 0 and one close to 1. This indicates that the classifier is relatively certain about the label of most models. We also see that the peak at 1 is significantly larger than the peak at 0. This reflects the imbalance of the training dataset which had approximately 2.5\% 0's.

There are, however, predictions other than just 0 and 1. This is because the classifier outputs some number $\hat{p}\in (0,1)$ which we can view as the classifier's prediction of likelihood of the true label being a 1. models with intermediary values represent inputs that the classifier is less certain about. Given this output, we can form a partition of the data into 0's and 1's by declaring that all outputs under some cut-off $p^{*}\in (0,1)$ should be labelled as 0, and the rest as 1. 

Typically, $p^{*} = \tfrac{1}{2}$, and indeed this was the value taken when assessing the accuracy of this classifier in the previous section.

Using $p^{*} = \tfrac{1}{2}$, one might assume that 96.7\% of the data that are predicted 0's have a true label of 0. This is not the case, however, because of the highly imbalanced nature of the dataset (2.5\% 0's and 97.5\% 1's). Using Bayes' theorem, and taking the accuracy to be 96.7\%, we see that
\begin{align}
    \begin{split}\label{eq:bayes}
        \mathbb{P}(TN \,|\, PN) &= \frac{\mathbb{P}(PN\,|\, TN) \mathbb{P}(TN) }{\mathbb{P}(PN)} \\
        &= \frac{0.967\times 0.025}{(0.967\times 0.025 + 0.033\times0.975)} \\
        &\approx 0.429
    \end{split}
\end{align}
Here, $PN$ is the event that a data point is `predicted negative', i.e. predicted to be 0, and $TN$ is the event that a data point is `true negative', i.e. that its true label is 0. It would appear then that, even though our classifier is very accurate, the imbalance of the dataset means that we cannot trust the set of predicted 0's to be a good representation of all $TN$ data.

We can improve our prediction by changing the cut-off value $p^{*}$. By changing this value, we can get a model which performs better under different performance metrics. Two metrics of interest to us are the 0-precision and the 0-recall.\footnote{Note that usually precision and recall are defined in reference to the 1's in the dataset. However, since the points of interest in our dataset were labelled with 0 we refer to these as 0-precision and 0-recall.} Both are a measure of how accurately the model can predict 0's. If $TN$ is the number of true negatives (i.e. the number of predicted 0's whose true label is 0) and $FN$ is the number of false 0's (i.e. the number of predicted 0's whose true value is 1), and similarly for TP, FP for true/false positive, then the 0-precision and 0-recall are defined as
\begin{align}
    \text{0-precision} &= \frac{TN}{TN + FN} \qquad
    \text{0-recall} = \frac{TN}{TN + FP}
\end{align}
In other words, the 0-precision tells us the probability that a data point has true label 0 given that it is predicted 0 (similarly to \eqref{eq:bayes})). The 0-recall, on the other hand, is the probability that we accurately predict a data point to be 0, given that its true label is 0. Typically, as we lower the value of $p^{*}$ we would increase the 0-precision at the cost of the 0-recall. Measuring these quantities on the full, unbalanced training data we can get a measure of the precision of the classifier on generic data.\footnote{This measure will likely be an over-estimation of the precision since the full unbalanced training dataset includes the data that the model was trained on and hence will perform slightly better than on data it has never seen before. Nonetheless, it is a helpful measure of the performance of our model.} A convenient way to capture this information is through the confusion matrix, which is a $2\times 2$ matrix given by
\begin{equation}
    \begin{array}{c|c}
        TP\ &\ TN \\\hline
        FP\ &\ FN
    \end{array}
\end{equation}

In our case, we can reduce the value of $p^{*}$ to raise the 0-precision significantly while maintaining the 0-recall sufficiently high to capture many models. We find that, with value of $p^{*} = 2.95 \times 10^{-6}$, applying the classifier to the full training dataset, we get the confusion matrix
\begin{equation}
    \begin{array}{c|c}
        TP = 531,531\ &\ TN = 3,369 \\\hline
        FP = 10,280\ &\ FN = 936
    \end{array}
\end{equation}
which gives a 0-precision and 0-recall of
\begin{equation}
    \text{0-precision} = 0.78257 \ , \qquad \text{0-recall} = 0.24683
\end{equation}

Applying this to the full unlabelled dataset, we predict a total of 214,837 models should have a label 0. Given the precision above, we might expect around 168,000 of them to pass the anomaly inflow test. Moreover, with the recall above, we can predict the total number of such models in the dataset to be around 681,000. By the nature of the anomaly inflow criteria, trivially combining any two of these models will lead to another which will pass the anomaly inflow criteria. The identified models therefore constitute a large set of building blocks, from which we can build a large number of theories inside the 6-dimensional string landscape.\footnote{Or at least which pass all known swampland criteria.} 

We can combine these results with the autoencoder built in previous sections. Plotting the 0's in red and the 1's in blue we find interesting clustering of the predicted 0's in the latent layer output. This is shown in Figure \ref{fig:first_classifier_output} below. This high clustering supports the claim that the auto-encoder has learnt complex properties of the 6-dimensional supergravity models that were not explicitly used in the training process. We can also use this clustering to guide our search for models in the landscape/swampland. The largest cluster of 0's is in LC-14 which may suggest that this cluster is particularly fertile for models inside the string-landscape. 

\begin{figure}[h!]
    \centering
    \includegraphics[width=0.75\linewidth]{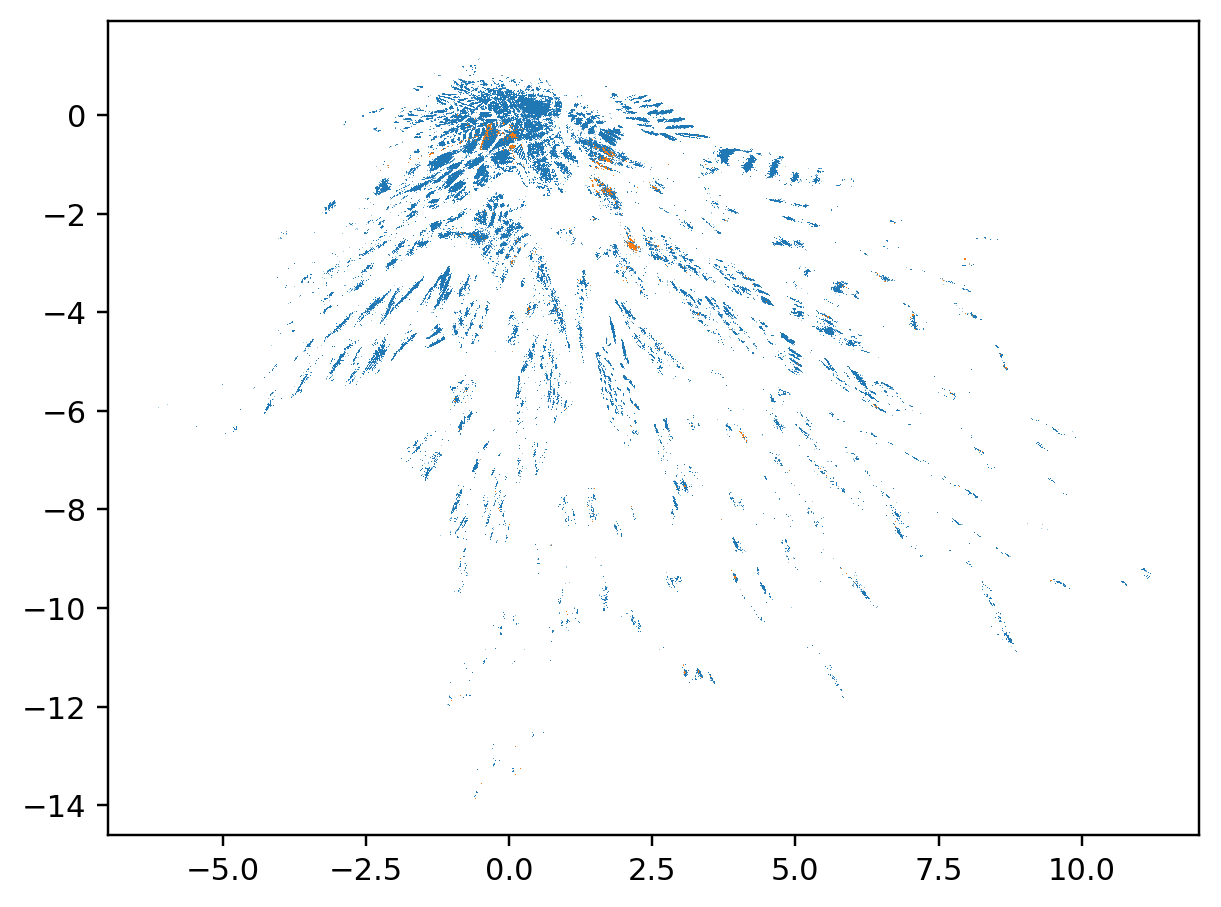}
    \caption{A plot of the output of the first classifier when fed through the encoder of the autoencoder. The predicted 0's are in red while the predicted 1's are in blue.}
    \label{fig:first_classifier_output}
\end{figure}

\subsubsection{The second classifier}\label{sec:second-classifier}

The second classifier was trained on dataset \ref{data2}. Recall that this dataset labelled with a 1 any model whose central charge satisfied
\begin{equation}
    c_{l} - \sum_{i} \frac{k_{i}\dim G_{i}}{k_{i} + \check{h}_{i}} < 0
\end{equation}
for all $Q\in \{ \Lambda_{S}\otimes \bbR \, | \, q,k_{0} \text{ fixed}, k_{i} \geq 0 \}$, where $(q,k_{0})$ are one of
\begin{equation}
    (q,k_{0}) = (-1,1),\, (0,0),\, (0,2), \, (1,1), \, (1,3), \, (2,0),\, (2,2), \, (2,4)
\end{equation}
If a model satisfies the condition above, it does not guarantee that it is in the swampland, as discussed in section \ref{sec:numerical_labelling}. However, it gives a strong indication that the model is inconsistent and can therefore guide our study of the swampland.

\begin{figure}[h!]
    \centering
    \includegraphics[width=0.75\linewidth]{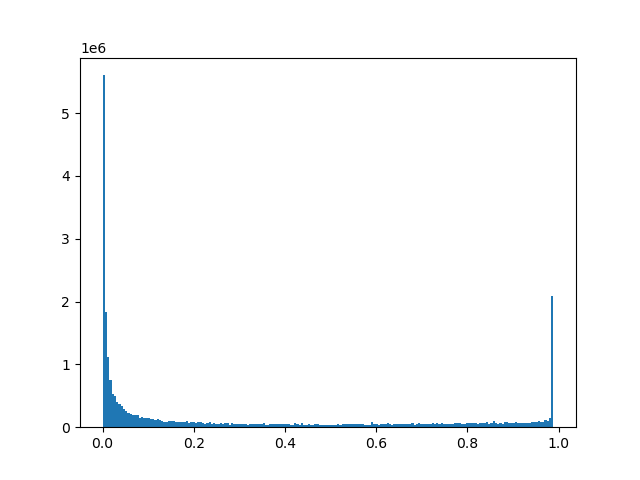}
    \caption{A frequency histogram of the outputs of the second classifier. The imbalance in the outputs reflects the known imbalance of the labelled dataset \ref{data2}, with around 22\% 1's.}
    \label{fig:classifier_hist_2}
\end{figure}

As before, we built a classifier NN with the same architecture as in Figure \ref{fig:classifier} and trained it on the labelled dataset \ref{data2}, down-sampled so that the number of 0's and 1's were equal. Once trained, this classifier had a validation accuracy of 85.5\%. Running the trained classifier on the full dataset, we found the histogram of outputs as in Figure \ref{fig:classifier_hist_2}.

As in the previous case, we can increase the precision (this time the 1-precision as the 1 labels are the data points of interest) by changing the value of $p^{*}$, at the cost of lowering the 1-recall. We find that by raising the value of $p^{*}$ to 0.989, we get the confusion matrix
\begin{equation}
    \begin{array}{c|c}
        TP = 8,315\ &\ TN = 87,572 \\\hline
        FP = 829\ &\ FN = 14,741
    \end{array}
\end{equation}
which gives a precision and recall of
\begin{equation}
    \text{1-precision} = 0.90933 \ , \qquad \text{1-recall} = 0.36064
\end{equation}

Applying this to the full dataset, we get a prediction of 1,909,359 models labeled 1 with around 1,736,000 models having a high likelihood of being inconsistent under anomaly-inflow.

\begin{figure}[h!]
    \centering
    \includegraphics[width=0.75\linewidth]{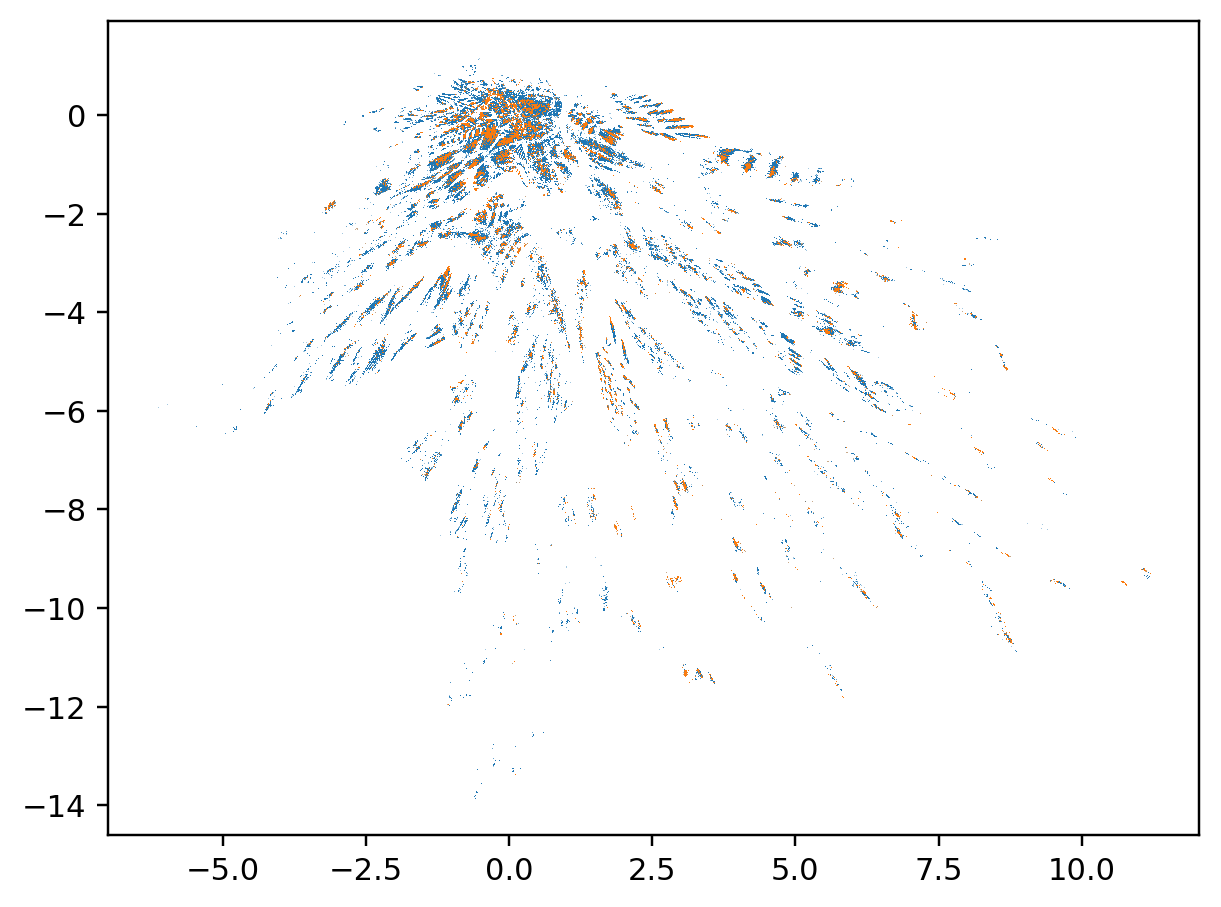}
    \caption{A plot of the output of the second classifier when fed through the encoder of the autoencoder. The predicted 0's are in red while the predicted 1's are in blue.}
    \label{fig:second_classifier_output}
\end{figure}

As before, we can plot these models in the latent layer of the autoencoder. This is done in Figure \ref{fig:second_classifier_output} where this time the 1's are plotted in red and the 0's are plotted in blue. This time, we see that the predicted 1's are much more spread out across the latent space. This may indicate that most local clusters contain models which are incompatible with the landscape. It would be interesting to examine the clusters with a relatively high density of predicted 1's to look for similarities and perhaps guide further swampland criteria. In contrast, there are some clusters with relatively few 1's, such as LC-06, LC-07, and LC-08. It might be interesting to examine these clusters to see if they contain models which are compatible with the landscape.

\section{Summary}\label{sec:summary}
Our work demonstrates how machine learning techniques may be used to study the string theory landscape from a bottom-up, swampland perspective. In section \ref{ML-Auto}, we built an autoencoder which we applied to the Gram matrices of anomaly-free/admissible models of $\mathcal{N} = (1,0)$ 6-dimensional supergravity. Not only did it exhibit interesting clustering in the latent layer, indicating that certain generic features of these models were identified by the algorithm, which we began to explore in \cite{ml-github-website}, but it was also able to identify models which were difficult to combine into totally anomaly-free theories. It was able to identify these features by just learning features of the Gram matrices of the irreducible components, and without using information around how models combine together or the constraint $\Delta + 29n_{T} \leq 273$.

We also built two classifiers in section \ref{sec:numerical_labelling} which label models with a high probability of either remaining consistent under probe brane insertion (i.e.~the first classifier \S~\ref{sec:first-classifier}) or becoming inconsistent under anomaly inflow conditions (i.e.~the second classifier \S~\ref{sec:second-classifier}). Checking for anomaly inflow is a non-trivial computation that would be too time consuming to perform using standard methods. However, the classifiers trained on suitably labelled datasets, were able to predict the likely (in)consistency of millions of models in a matter of minutes. This shows the ability of machine learning algorithms to learn highly complex properties of supergravity theories. When both the classifier and autoencoder results are used in tandem we found that models which passed the anomaly inflow conditions appeared to cluster together, v.s.~Figure \ref{fig:classifier_1}. We identified one cluster, LC-14, that contained the largest concentration of consistent models. It would be interesting to investigate this subset of models in particular.

It is clear from the autoencoder's clustering of models, discussed in subsection \ref{sec:analysis}, that the space of admissible/anomaly-free models share similar features encoded within their Gram matrices. Each enumerated cluster then allows one to work with subsets of the data with similar properties, rather than the full dataset. This is an alternative way of working with the data that can give one access to a unique set of models with very similar gram matrices. Such collections of data would be highly useful for more targeted investigations of the swampland/landscape. We provide the model data (i.e. \(n_T\), \(\lambda_+, \lambda_-\), and associated hypermultiplet content) for each cluster in separate files which can be found here \cite{ml-github-repo} for the readers to use. It would be interesting to explore the properties of these clusters in more detail, which may help to uncover new properties of supergravity theories of interest.

In terms of the machine learning techniques applied, there are two major directions in which we could improve this work.  It was noted that both our autoencoder and classifier models incorporated only data about a solution's Gram matrix, ignoring more refined data about the gauge groups and representation data. One easy solution could be to include the rank of the simple gauge groups as an extra parameter in the input. Alternatively, our early prototypes for both models incorporated a mix of a model's group representation vector and Gram matrix data, but ultimately we found that this did not improve performance over the Gram matrix data alone.  The cause of this may be that specifying the full representation content of a solution requires a vector with over 15,000 entries, and it may be that the relevant information sat too sparsely in that space to generate a useful signal compared to the Gram matrix.  A possible fix for this would be to use a multi-network solution where one network distils the representation data into a useful lower-dimensional form which could then be fed into a second network.  Another tactic we may apply in the future is the use of a sparse autoencoder \cite{Sparse} (as opposed to the traditional one we employed) to force the unsupervised classification to occupy fewer clusters. The ability to fine-tune the resulting spread of data by adding a sparsity constraint to the loss may lead to a more easily digestible set of clusters. These refinements of the input data may also help to increase the accuracy of the classifiers, helping to refine the search for the landscape/swampland.

\appendix
\section{An ambiguity when \texorpdfstring{$\lambda_{+}(\bbG) = 0$}{lambda{+}(G)=0} and \texorpdfstring{$\lambda_{-}(\bbG) < n_{T}$}{lambda{-}(G)<nT}}\label{app:A}

As mentioned previously, there is an ambiguity in the definition of the anomaly coefficients $a,b_{i}$ when $\lambda_{+}(\bbG) = 0$ and $\lambda_{-}(\bbG) < n_{T}$. To see this, note that we can always shift $a$ and $b_{i}$ by some constant vector $x$ provided
\begin{equation}
    a\cdot x = b_{i}\cdot x = x\cdot x = 0
\end{equation}
If such an $x$ exists then the anomaly factorisation \eqref{eq:anomaly_factorisation} can be written in terms of $a,b_{i}$ or $\tilde{a} = a+x$ and $\tilde{b}_{i} = b_{i}+x$. It is not hard to check that such an $x$ exists only if $\lambda_{+}(\bbG) = 0$ and $\lambda_{-}(\bbG)<n_{T}$.

At the level of the bulk theory, such a choice is not of particularly great importance. However, when a probe brane is inserted, the anomaly inflow to the string depends directly on the chosen values of $a,b_{i}$. Since this can affect the central charges and flavour current levels on the probe string, such a choice can change the conclusion when evaluating the consistency when coupled to probe strings, as described at the end of section \ref{sec:consistency_review}. Without good physical selection criteria for the value of $a,b_{i}$, the consistency of the model is ambiguous.

To illustrate this fact, let's consider a very simple model with a single simple gauge group $\SU{5}$ and a single adjoint hypermultiplet making up the non-trivial matter. In the notation of \eqref{eq:example_1} we have
\begin{equation}
    \SU{5}\, : \ \textbf{24}
\end{equation}
It is easy to check that when $n_{T} = 9$ all anomalies (gauge, gravitational, and mixed) vanish. Hence, the Gram matrix $\bbG$ is the 0 matrix. Clearly this has $\lambda_{+}(\bbG) = 0$ and $\lambda_{-}(\bbG) = 0 <n_{T} = 9$. It is also clearly factorisable, but to ensure that we satisfy the ghost-free condition we cannot set $a,b$ to 0. Two possible choices for the matrix $\bbD = (a \ b)$, and a value for the tensor multiplet VEVs $j$ which results in a ghost free theory in both cases, are
\begin{align}
    \bbD = \left( \begin{array}{cccc}
        3 & -1 & ... & -1 \\
        3 & -1 & ... & -1
    \end{array} \right) \qquad
    \tilde{\bbD} = \left( \begin{array}{cccc}
        3 & -1 & ... & -1 \\
        9 & -3 & ... & -3
    \end{array} \right) \qquad j = \left( \begin{array}{ccccc}
        2 & -1 & 0 & ... & 0
    \end{array} \right).
\end{align}
Note that both choices of $\bbD$ satisfy all but the probe brane consistency requirements, including being able to be embedded into the odd rank 10 charge lattice $\text{I}_{(1,9)}$.

To check the consistency when coupled to probe strings, we consider inserting a string with charge
\begin{equation}
    Q = \left( \begin{array}{ccccc}
        0 & 1 & 0 & ... & 0
    \end{array} \right).
\end{equation}
Note that this value of $Q$ gives
\begin{equation}
    c_{r} = 0 \ , \quad c_{l} = 8 \ , \quad k_{l} = 0 \ , \quad k = Q\cdot b = 1 \ , \quad \tilde{k} = Q\cdot \tilde{b} = 3 \ , \quad Q\cdot j = 1
\end{equation}
and hence satisfies the conditions \eqref{eq:inflow_cond_1} for both $\bbD$ and $\tilde{\bbD}$. We then consider the inflow consistency condition \eqref{eq:inflow_unitarity} for both $\bbD$ and $\tilde{\bbD}$. in the first case, we have
\begin{equation}
    \bbD\, : \qquad \frac{k\dim \SU{5}}{k + 5} = \frac{24}{6} = 4 < c_{l}
\end{equation}
and hence passes the inflow criterion at this point. In the second case, however, we find
\begin{equation}
    \tilde{\bbD} \, : \qquad \frac{\tilde{k} \dim\SU{5}}{\tilde{k} + 5} = \frac{72}{8} = 9 > c_{l}
\end{equation}
and hence we would determine that this model is inconsistent with probe branes.

We see then that the consistency of such models is ambiguous. For this reason, we have excluded these cases from both of our labelling processes. In each case we labelled such models opposite to not possess whichever characteristic we were looking for, i.e. labelled them 1 in the first labelling process and 0 in the second.

\section{Acknowledgements}
We would like to thank Anthony Ashmore, Katrin Becker, Gabriel Larios, Ergin Sezgin, and Daniel Waldram for interesting discussions and useful comments. Portions of this research were conducted using the advanced computing resources provided by Texas A\&M High Performance Research Computing. NB and DT were supported by the NSF grant PHY-2112859.

\bibliographystyle{JHEP}
\bibliography{references.bib}

\end{document}